\documentclass[a4paper,fleqn,11pt]{article}

\usepackage[margin=1in]{geometry}
\usepackage[authoryear,longnamesfirst]{natbib}
\usepackage{capt-of}
\usepackage{graphicx}
\usepackage{amsmath,amssymb,mathtools,bm}
\usepackage{booktabs}
\usepackage{placeins}
\usepackage[svgnames,dvipsnames]{xcolor}
\usepackage[colorlinks,citecolor=NavyBlue,linkcolor=NavyBlue,urlcolor=NavyBlue]{hyperref}
\usepackage{orcidlink}

\providecommand{\sep}{\unskip, }
\newenvironment{keywords}
  {\small\par\noindent\textbf{Keywords: }}
  {\par}

\newcommand{\tp}{\tilde{p}}
\newcommand{\pstar}{p^{\star}}
\newcommand{\xstar}{x^{\star}}
\newcommand{\dstar}{\delta^{\star}}
\newcommand{\sct}[1]{\mathrm{#1}}
\newcommand{\fXi}{f_{\Xi}}
\newcommand{\FXi}{F_{\Xi}}
\newcommand{\fsig}{f_{\sigma}}
\newcommand{\Expect}{\mathbb{E}}
\newcommand{\smallbraceunder}[2]{%
  \underset{#2}{\scalebox{0.7}{$\underbrace{\scalebox{1.4286}{$\displaystyle #1$}}$}}}
\newcommand{\xvec}{\bm{x}}
\newcommand{\xivec}{\bm{\xi}}
\newcommand{\muvec}{\bm{\mu}}
\newcommand{\xipar}{\xi_{\parallel}}
\newcommand{\xiperp}{\xi_{\perp}}
\newcommand{\epar}{\bm{e}_{\parallel,\ell}}
\newcommand{\eperp}{\bm{e}_{\perp,\ell}}
\newcommand{\cpar}{c_{\parallel}}
\newcommand{\cperp}{c_{\perp}}
\newcommand{\eu}{\bm{e}_{u}}
\newcommand{\xbar}{\bar{\xvec}}
\newcommand{\tpr}{\hat{\tp}}
\newcommand{\pstarr}{\hat{p}^{\star}}
\newcommand{\pr}{\hat{p}}

\title{Mollified--sharp decomposition: a probabilistic regularization of parametric POD for shock-bearing flows}

\author{Oliver T. Schmidt~\orcidlink{0000-0002-7097-0235}\\
  University of California San Diego, La Jolla, CA 92093, USA\\
  \texttt{oschmidt@ucsd.edu}}

\date{}

\begin{document}
\let\WriteBookmarks\relax
\def\floatpagepagefraction{1}
\def\textpagefraction{.001}

\shortcites{catalani2023comparative}

\maketitle

\begin{abstract}
This paper introduces the mollified--sharp decomposition, a probabilistic
regularization of moving shocks in parametric reduced-order models. Each
detected shock location is treated as a random variable with a prescribed
probability density. Averaging over this artificial distribution replaces the
localized pressure change by a smooth transition whose spatial extent is set
by the density width $\sigma$, rather than by direct filtering of the pressure
field. Each snapshot is decomposed exactly into a
regularized mollified field and a local sharp correction that restores the
shock. This probabilistic construction and exact additive split define the
general method; the detector, kernel, treatment of multiple shocks, alignment
coordinates, and regression are implementation choices. In the present
realization, a calibrated indicator detects shocks, a compactly supported
Wendland kernel mollifies them, and a peak-normalized and, where necessary,
partitioned weight derived from each shock-location probability density defines
the centroid, principal axes, and scales of its local alignment domain. Separate
POD--GPR models represent the mollified field and aligned corrections, with
additional regressions for shock presence and alignment. The method is
demonstrated on the transonic airfoil pressure data of
\citet{catalani2023comparative} and compared with a standard POD--GPR model
constructed from the same data and POD-energy criterion. The mollified--sharp
model reduces the mean test-set relative $L^2$ pressure error by 31.2\% and the
mean test-set surface-pressure-coefficient error by 33.2\%, while also
improving the predicted shock locations and pressure changes; the trade-off is
a median online evaluation time 2.24 times as long. The construction is
applicable in principle to other
parameter-dependent fields with moving sharp features, such as moving
material interfaces in multiphase flows.

\begin{center}
{Distribution A: Approved for public release; distribution is
unlimited. AFRL-2026-3564.}
\end{center}
\end{abstract}


\section{Introduction}
\label{sec:intro}

Shocks in compressible flows pose a major difficulty for parametric proper
orthogonal decomposition (POD) \citep{sirovich1987turbulence,berkooz1993proper}
and for linear model order reduction more generally
\citep{lucia2004reduced,benner2015survey}, because the flow field depends
nonsmoothly on the parameters. A moving shock is poorly represented by a fixed
linear basis: in POD, the modes superpose snapshots whose shocks occur at
different locations. A truncated expansion may therefore exhibit oscillations
or staircase artifacts associated with shock locations represented in the
training snapshots \citep{gottlieb1997gibbs} and cannot, in general, recover a
sharp jump at an unsampled target location. This behavior is reflected in the slow
Kolmogorov $n$-width decay of transport-dominated solution sets
\citep{ohlberger2016reduced,greif2019decay}, which can necessitate large bases
and dense training data. The associated cost is especially important in
multidisciplinary design optimization (MDO), where coupled discipline models
must be evaluated and differentiated many times. Although scalable coupling
and derivative methods reduce part of this cost, practical many-query design
still depends on accurate and inexpensive reduced-order flow-field models
\citep{HwangMartins2018MAUD,HwangMartins2018Surrogate}.

We consider nonintrusive reduced-order models for steady aerodynamic fields.
POD represents each field in a low-dimensional basis, and regression evaluates
the modal coefficients over the parameter space
\citep{buithanh2004aerodynamic,zimmermann2012improved,franz2014interpolation,yondo2018review}.
Here the coefficients are interpolated by Gaussian-process regression (GPR),
giving the POD--GPR model used throughout
\citep{guo2018reduced,guo2019data}. Although this approach is simple and
efficient away from shocks, in transonic flow it often produces smeared or
staircase-like shocks and oscillatory surface-pressure distributions. Denser
sampling of the shock-sensitive part of the parameter space only partly
alleviates the problem. Weighted POD can adapt the influence of available
snapshots to the current parameter point
\citep{vanschie2025weightedproperorthogonaldecomposition}, while local POD
models can separate distinct flow regimes \citep{dupuis2018surrogate};
neither approach directly removes the difficulty of representing a shock that
moves within the region emphasized by the basis.

Several methods address this transport barrier directly
\citep{peherstorfer2022breaking,hesthaven2026nonlinear}. Local or online bases
adapt the approximation space to the parameter or state
\citep{amsallem2012nonlinear,peherstorfer2020model}, whereas transformation
methods align moving features before applying a linear decomposition. Examples
of the latter include symmetry reduction \citep{rowley2000reconstruction},
shifted POD \citep{reiss2018shifted}, transformed snapshots
\citep{welper2017interpolation}, transported snapshots for steady flows with
moving shocks \citep{nair2019transported}, registration
\citep{taddei2020registration,ferrero2022registration}, and implicit feature
tracking \citep{mirhoseini2023model,zucatti2026model}. These methods can greatly
improve low-rank approximation, but a single global transformation may be
difficult to define when several shocks move independently, meet boundaries,
or appear and disappear, as commonly occurs across a transonic flight envelope
\citep{razavi2025registration}.

Optimal transport offers a related description of moving features and has led
to advection modes \citep{iollo2014advection}, reduction in Wasserstein spaces
\citep{ehrlacher2020nonlinear}, and displacement interpolation
\citep{cucchiara2024model}. It provides a feasible route for interpolating
shock-location probability densities or constructing nonlinear maps between
full fields. In the present setting, however, transporting the shock-location
probability density would not by itself determine the signed pressure
correction when several shocks differ in amplitude and geometry or appear and
disappear. We instead use the probability density for regularization and
represent each sharp correction in a local aligned frame. Optimal-transport
formulations remain a promising alternative for future extensions.

This paper addresses moving shocks through the \emph{mollified--sharp decomposition},
$p=\widetilde{p}+p^{\star}$. For each detected shock, we associate its
deterministic location with an artificial probability distribution centered
on the detected position, solely as a means of regularizing the shock-induced
sharp transition. Taking the expectation over this distribution spreads the
sharp transition over a controlled width. The resulting mollified snapshots
are better suited to low-rank POD representation. This construction connects
classical mollification with snapshot data
\citep{friedrichs1944identity}. A shock indicator locates the front
\citep{jameson1981numerical,ducros1999large,persson2006subcell}, and a kernel of
width $\sigma$ forms the empirical shock-location probability density
\citep{silverman1986density}. The formulation does not require the particular
indicator or kernel used here and, because it is applied only to completed
snapshots, does not alter the full-order solver. Its smoothing effect is
analogous to artificial viscosity
\citep{vonneumann1950method,guermond2011entropy} and to the smearing of mean
fields caused by uncertain shock locations
\citep{lin2006predicting,poette2009uncertainty}; here, however, smoothing is
introduced deliberately as a data transformation.

The mollified field $\widetilde{p}$ is smooth across each detected shock and is
represented by one global POD--GPR model, whereas the exact residual
$p^{\star}=p-\widetilde{p}$ restores the sharp local structure. To
accommodate multiple shocks, we assign them to fixed geometric sectors and
express each sharp component in a local frame that removes its translation,
rotation, and scale before forming a local POD--GPR model. Separate GPR models
provide the frame parameters and determine whether each component is present. Treating the
components separately avoids a single global deformation and remains well
defined when shocks appear or disappear. The decomposition is exact on every
training snapshot, and the online reconstruction is explicit. The aligned
residual shares the explicit treatment of shock position used in shock fitting
and tracking
\citep{moretti1987computation,zahr2018optimization} and builds on the earlier
observation that separating the shock region improves reduced models
\citep{lucia2003reduced}.

The method is assessed using transonic pressure fields from the RAE2822
Reynolds-averaged Navier--Stokes (RANS) database of
\citet{catalani2023comparative}. We consider both the full pressure field and
its surface distribution. The decomposition itself is not specific to pressure
and can be applied to other flow variables with shock-induced jumps. We compare
the method with a standard POD--GPR model constructed from the same data.
Section~2 develops the probabilistic mollification, and Section~3 defines and
aligns the local sharp correction. Section~4 describes the database and
constructs the parametric reduced-order model. Section~5 presents the results,
while Sections~6 and~7 discuss the method and summarize the conclusions.
Database-specific implementation details are given in the appendices.

\FloatBarrier
\section{Probabilistic mollification}
\label{sec:moll}

The central idea of this paper is to split the pressure into a mollified part
and a local sharp correction:
\begin{equation}
  p =
  \smallbraceunder{\tp}{\substack{\text{mollified}\\[-1pt]\text{part}}}
  +
  \smallbraceunder{\pstar}{\substack{\text{local sharp}\\[-1pt]\text{correction}}},
  \label{eq:decomp}
\end{equation}
We call this the \emph{mollified--sharp decomposition}. To construct $\tp$, the
deterministic shock location is associated, even in the absence of uncertainty,
with an artificial probability distribution solely as a means of regularizing
the shock-induced sharp transition. Averaging over this distribution connects
the construction to classical mollification. Its width is controlled by the
user through the smoothing-kernel parameter $\sigma$, which sets the
regularization scale. The resulting mollified part is smooth across each
detected shock and is therefore better suited to low-rank POD approximation and
regression over the parameter space. The local sharp correction
$\pstar=p-\tp$ restores the sharp structure. As in
\citet{catalani2023comparative}, we focus on pressure, although the decomposition
can in principle be applied to other flow variables with shock-induced jumps.

Subsection~\ref{sec:moll:idealized} first introduces an idealized
one-dimensional problem to demonstrate the construction and its connection to
classical mollification. Subsection~\ref{sec:moll:empirical} then constructs
the shock-location distribution and the mollified field empirically from the
RAE2822 pressure data of \citet{catalani2023comparative}.
Section~\ref{sec:sharp} defines and aligns the local sharp correction, and
Section~\ref{sec:surrogate} constructs the parametric reduced-order model.

\subsection{The idealized problem}
\label{sec:moll:idealized}

Consider a one-dimensional idealization of a pressure profile containing a
single shock. The pressure is constant on either side of the shock and changes
by $\Delta p$ at the known location $\xstar$:
\begin{equation}
  p(x; \xstar) \;=\; p(x_-) \;+\; \Delta p \cdot H(x - \xstar),
  \qquad
  \Delta p \;\equiv\; p(x_+) - p(x_-),
  \label{eq:1d_p}
\end{equation}
Here, $H$ is the Heaviside step function, and the points $x_-$ and $x_+$ lie
upstream and downstream of the shock, respectively, such that
$x_-<\xstar<x_+$. The pressure change is positive, $\Delta p>0$, for a
compression shock. The classical derivative is not defined at the jump.
However, in the sense of distributions, the derivative of the Heaviside
function is a Dirac delta. Differentiating Eq.~\eqref{eq:1d_p} therefore
represents the entire pressure change as a Dirac delta of weight $\Delta p$:
\begin{equation}
  \frac{\partial p}{\partial x}(x; \xstar) \;=\; \Delta p \cdot \delta(x - \xstar),
  \label{eq:1d_grad}
\end{equation}
The interval from $x_-$ to $x_+$ contains the shock location. Because a unit
Dirac delta integrates to one over any interval containing its location,
integrating the pressure gradient over this interval recovers the full pressure
change:
\begin{equation}
  \int_{x_-}^{x_+}\!\frac{\partial p}{\partial x}(x;\xstar)\,dx \;=\; p(x_+) - p(x_-) \;=\; \Delta p.
  \label{eq:ftc}
\end{equation}
We now promote the known, deterministic shock location to a random variable
$\Xi$ with probability density function (PDF) $\fXi$. The PDF is nonnegative,
has unit integral, and distributes artificial probability mass over the
possible shock coordinates. Its cumulative distribution function (CDF) is
\begin{equation}
  \FXi(x) \;=\; \int_{-\infty}^{x}\!\fXi(x')\,dx',
  \label{eq:FXi}
\end{equation}
so that $\FXi(x)$ is the probability that the shock lies at or upstream of
$x$. It increases from zero to one as $x$ passes through the support of the
density. For commonly used symmetric, single-peaked choices, such as Gaussian
and Wendland kernels, the density is centered at the deterministic shock
coordinate. This coordinate is also the mode, or location of maximum
probability density,
$\xstar=\operatorname*{arg\,max}_{x}\fXi(x)$. We define the mollified pressure
as the expectation over the promoted shock location:
\begin{equation}
  \tp(x) \;\equiv\; \Expect_{\Xi \sim \fXi}\!\bigl\{p(x;\Xi)\bigr\}.
  \label{eq:1d_p_tilde_def}
\end{equation}
This expectation averages pressure profiles whose shock positions are shifted
according to $\fXi$. For a fixed coordinate $x$, $H(x-\Xi)$ equals one when
$\Xi\leq x$ and zero otherwise. Its expectation is therefore the probability
that the shock lies at or upstream of $x$, which is exactly $\FXi(x)$.
Similarly, averaging $\delta(x-\Xi)$ over all possible shock positions returns
the probability density at $x$. These results follow directly from the
definitions of expectation, PDF, and CDF:
\begin{equation}
  \Expect_{\Xi \sim \fXi}\!\bigl\{H(x-\Xi)\bigr\} \;=\; \FXi(x),
  \qquad
  \Expect_{\Xi \sim \fXi}\!\bigl\{\delta(x-\Xi)\bigr\} \;=\; \fXi(x),
  \label{eq:H_to_F}
\end{equation}
Substituting the first identity into Eq.~\eqref{eq:1d_p} replaces the
Heaviside step by a CDF-shaped ramp:
\begin{equation}
  \tp(x) \;=\; p(x_-) \;+\; \Delta p \cdot \FXi(x),
  \label{eq:1d_p_tilde}
\end{equation}
Applying the second identity to Eq.~\eqref{eq:1d_grad} gives the corresponding
mollified pressure gradient:
\begin{equation}
  \frac{\partial \tp}{\partial x}(x) \;=\; \Delta p \cdot \fXi(x).
  \label{eq:1d_p_tilde_grad}
\end{equation}
In the far-upstream limit, $\FXi\to0$, and the mollified pressure tends to
$p(x_-)$. In the far-downstream limit, $\FXi\to1$, and the mollified pressure
tends to $p(x_-)+\Delta p=p(x_+)$. The mollification therefore preserves the
total pressure change while spreading it over the characteristic width of the
CDF transition. At the gradient level, the singular Dirac delta is replaced by the
PDF-shaped profile $\fXi$. Because $\fXi$ has unit integral, the area under the
mollified pressure gradient remains $\Delta p$.

Subsection~\ref{sec:moll:empirical} transfers this construction to the pressure
data. There, an empirical shock-location probability density supplies the PDF-shaped shock
contribution to a modified pressure gradient, and a data-derived estimate of
the pressure change sets its amplitude. Integrating that modified gradient then
produces the mollified pressure field.

\FloatBarrier
\subsection{Empirical construction}
\label{sec:moll:empirical}

\begin{figure}[!t]
  \centering
  \includegraphics[width=\textwidth]{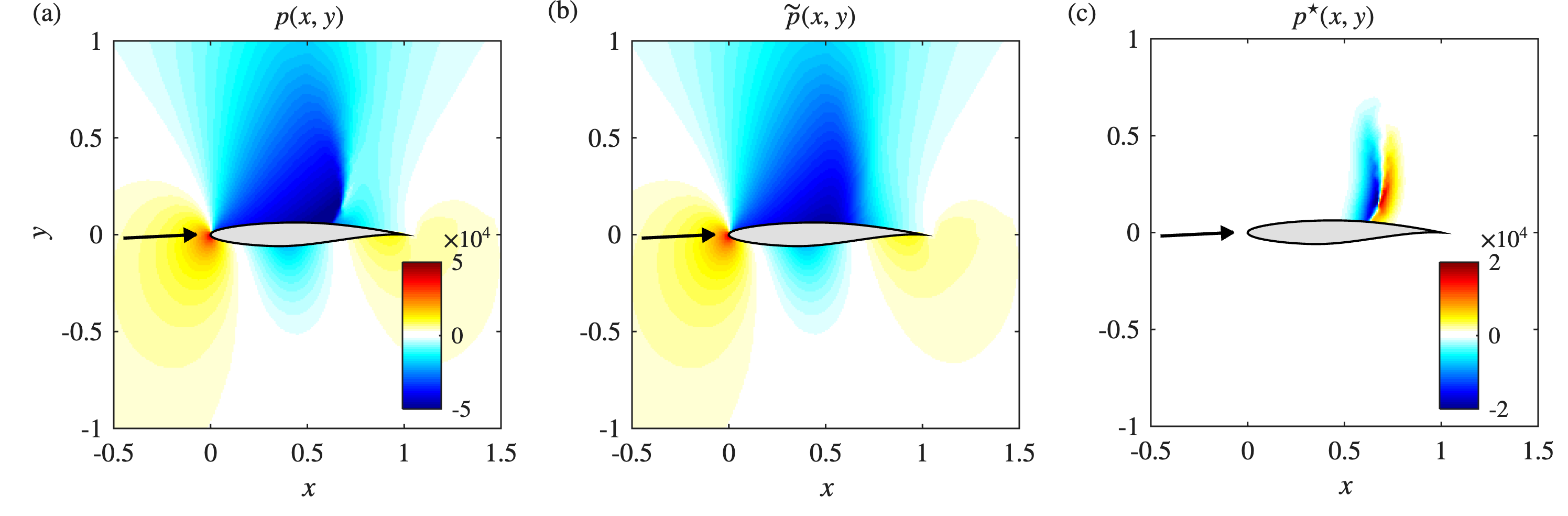}\\[-4.5mm]
  \includegraphics[width=\textwidth]{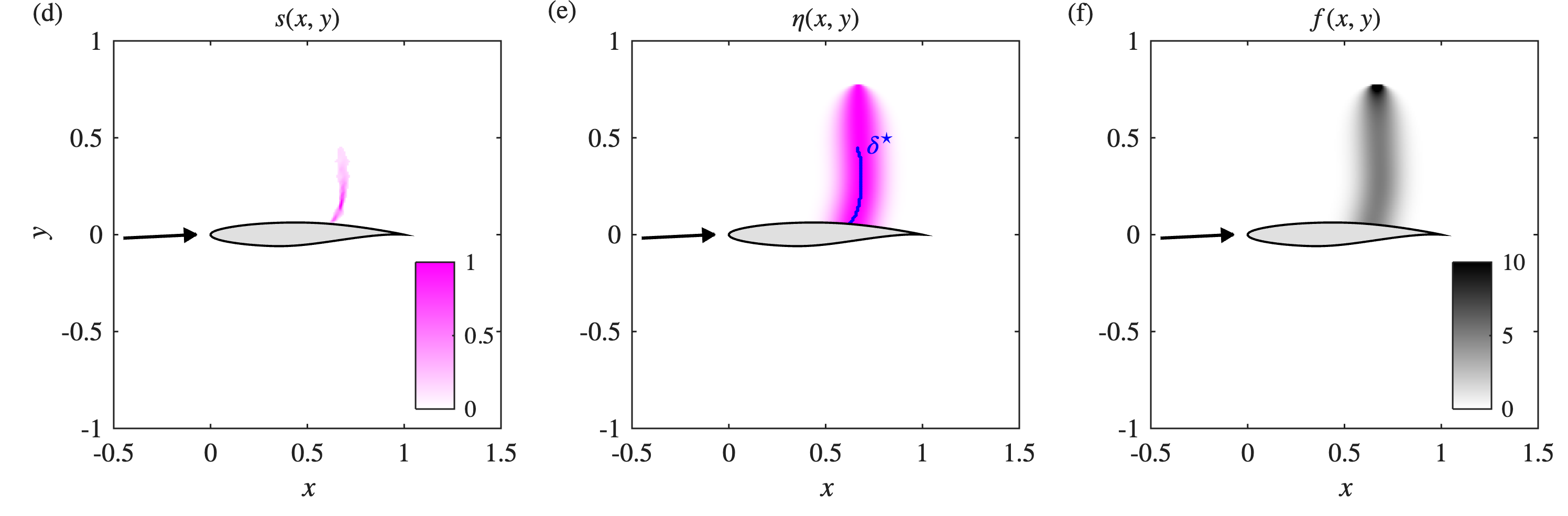}
  \caption{Overview of the empirical mollified--sharp decomposition
  $p=\tilde{p}+p^{\star}$ for a representative transonic RAE2822 pressure
  snapshot at $\alpha=2.78^{\circ}$ and $M_\infty=0.792$. The original
  pressure field $p(x,y)$ (a) is separated into the mollified part
  $\tilde{p}(x,y)$ (b) and the local sharp correction $p^{\star}(x,y)$ (c).
  The shock indicator $s(x,y)\in[0,1]$ (d) identifies the shock front. In
  panel (e), the blue curve is the shock-line trace $\delta^{\star}$ extracted
  from the indicator, and the magenta field is the mollification weight
  $\eta(x,y)\in[0,1]$. Panel (f) shows the empirical shock-location PDF
  $f=f_{\sigma}\ast\delta^{\star}$. For the single shock shown here, $\eta$
  is the peak-normalized counterpart of $f$: the two fields have the same
  shape, but $\eta$ has unit peak whereas $f$ has unit integral along each
  fixed-$y$ line. The factor $(1-\eta)$ weights the smooth-background
  contribution to the modified pressure gradient.}
  \label{fig:decomposition-example}
\end{figure}

The CFD fields are first interpolated to the fixed Cartesian grid
$(x_i,y_j)$ used throughout, as detailed in Appendix~\ref{app:algorithm}. All
spatial coordinates and kernel widths are nondimensionalized by the airfoil
chord. The idealized construction is then applied independently along the
streamwise coordinate $x$ at each fixed Cartesian $y$ coordinate of a pressure
snapshot. Fig.~\ref{fig:decomposition-example} previews the complete construction on a
representative snapshot from the RAE2822 database; the database and retained
parameter samples are described in Section~\ref{sec:database}. The
example is introduced here only to provide a visual overview. The upper row
shows the original pressure $p$, the mollified pressure $\tilde{p}$, and their
difference $p^{\star}=p-\tilde{p}$. The mollified field is smoother in the
shock neighborhood, while $p^{\star}$ contains the local sharp correction
needed to recover the original field. The lower row previews the empirical
construction of $\tilde{p}$. The indicator $s$ identifies the shock front,
from which the shock-line trace $\delta^{\star}$ is extracted. Convolution of
this trace with the smoothing kernel $f_{\sigma}$ distributes its mass over a
width controlled by $\sigma$. Normalization along each fixed-$y$ line produces
the unit-area shock-location PDF $f$, and peak normalization gives the
dimensionless mollification weight $\eta$. These quantities are defined below,
and the choice of $\sigma$ is discussed at the end of this section.

Shock fronts are located by an indicator $s\in[0,1]$ that is near one along a
detected front and near zero away from it. Many indicator definitions are
available \citep{jameson1981numerical,ducros1999large,persson2006subcell}. The
formulation can accommodate different shock indicators; the one used for the
present results is specified in Appendix~\ref{app:detection}. The construction
is also not restricted to compression shocks. The sign of the pressure change
is retained, so any sufficiently localized sharp transition identified by the
chosen detector can be treated in the same way.

Let $\ell$ index a sharp-correction component. In the present implementation,
each component is associated with one of the fixed sectors introduced in
Section~\ref{sec:sharp}. For a component present on a fixed-$y$ line, let
$\xstar_{\ell}(y)$ denote its streamwise shock location. The locations
extracted from $s$ form a shock-line trace. In the present implementation, this
trace is represented by the graph delta
$\dstar_{\ell}$ defined in Appendix~\ref{app:graph-delta}. Collapsing the
finite-width indicator to a graph is not intrinsic to the mollification. It is
used here to remove variations in the apparent detector thickness and thereby
make the prescribed kernel width $\sigma$ the dominant smoothing scale. If a
detector already produced a sufficiently consistent finite-width field, its
sector-restricted response could instead be normalized along each fixed-$y$
line to define $f_{\ell}$ directly, with $\eta_{\ell}$ obtained by peak
normalization. More generally, another nonnegative, unit-mass representation
of the detected front could replace $\dstar_{\ell}$. For the graph-delta route
used here, $\ast$ denotes spatial convolution with the two-dimensional
smoothing kernel $\fsig$ of width $\sigma$ on the Cartesian grid. For an
isolated component, the convolution gives the empirical shock-location
probability density $f_{\ell}$ and its CDF $F_{\ell}$:
\begin{equation}
  f_{\ell} \;=\; \fsig \ast \dstar_{\ell},
  \qquad
  F_{\ell}(x,y) \;=\; \int_{-\infty}^{x} f_{\ell}(x',y)\,dx'.
  \label{eq:fl}
\end{equation}
The two-dimensional kernel has unit total mass, whereas the graph trace carries
one unit of streamwise mass on every active fixed-$y$ line. After linewise
normalization, including boundary correction where the support intersects the
airfoil or outer domain, $f_{\ell}(\,\cdot\,,y)$ has unit integral in $x$.
Thus, for each fixed $y$, it is a one-dimensional probability density in the
streamwise coordinate, not a joint density in $(x,y)$. For a front normal to
the streamwise direction, this density has width $\sigma$ and is centered at
$\xstar_{\ell}(y)$. With $y$ treated as a parameter, it is the empirical
realization of the artificial random shock location $\Xi$ from
Section~\ref{sec:moll:idealized}, with the identification
\begin{equation}
  \fXi \;\equiv\; f_{\ell},
  \qquad \FXi \;\equiv\; F_{\ell},
  \label{eq:postulate}
\end{equation}
which connects the probabilistic idealization to the data-based construction.
If the front tangent makes an angle $\theta$ with the vertical direction
normal to the streamwise sampling line, its center remains
$\xstar_{\ell}(y)$, but its effective streamwise width becomes
$\sigma/\cos\theta$. The grid-level construction is detailed in
Appendix~\ref{app:graph-delta}.

The probability density $f_{\ell}$ determines where the shock contribution is distributed.
As made explicit by the first term of Eq.~\eqref{eq:mod_grad}, a separate
dimensionless weight is needed to suppress the raw pressure gradient in the
same shock neighborhood. We obtain this mollification weight by normalizing
$f_{\ell}$ to unit peak on each fixed-$y$ line. For isolated or nonoverlapping
components, this gives the first relation below directly. Where component
supports overlap, the reported implementation first applies this relation to
obtain preliminary weights, partitions them proportionally so that their sum
does not exceed one, and then normalizes each partitioned weight in $x$ to
obtain the final linewise probability density. The same symbols
$\eta_{\ell}$ and $f_{\ell}$ are retained for these final fields; the discrete
construction is given in Appendix~\ref{app:algorithm}. Thus,
\begin{equation}
  \eta_{\ell}(\xvec) \;=\; \frac{f_{\ell}(\xvec)}{\max_{x'} f_{\ell}(x',y)},
  \qquad
  \eta(\xvec) \;=\; \sum_{\ell} \eta_{\ell}(\xvec) \;\in\; [0,1].
  \label{eq:eta}
\end{equation}
For an isolated component, $\eta_{\ell}$ has unit peak rather than unit area.
For an untruncated shock normal to the streamwise line, $\eta_{\ell}$ and
$f_{\ell}$ therefore have the same support and shape, as illustrated in panels
(e) and (f) of Fig.~\ref{fig:decomposition-example}.

The probability density fixes the shape and location of the mollified shock contribution
but not its signed amplitude. We estimate that amplitude by integrating the
raw pressure gradient over the support
$\operatorname{supp}\eta_{\ell}=[x_-,x_+]$, weighted by
$\eta_{\ell}$:
\begin{equation}
  \Delta p_{\ell} \;=\; \int_{x_-}^{x_+}\!\eta_{\ell}(x)\,\frac{\partial p}{\partial x}(x)\,dx.
  \label{eq:jump}
\end{equation}
Eq.~\eqref{eq:jump} retains the sign of the pressure change: it is positive
for a compression shock and negative for a localized pressure decrease. In the
discrete implementation it is evaluated independently along $x$ at each fixed
$y$, so the amplitude is line dependent, $\Delta p_{\ell}(y)$. The argument $y$ is
suppressed below when no ambiguity arises. Eq.~\eqref{eq:jump} is the
weighted, data-based counterpart of Eq.~\eqref{eq:ftc}. In the idealized case,
the entire pressure gradient is concentrated at the jump, where the
unit-peak weight equals one, and the two expressions coincide. For a resolved
CFD profile, the weight instead isolates the part of the local pressure change
associated with the detected shock.

Real CFD profiles also contain smooth pressure variation away from the shock,
which was absent from the piecewise-constant model of
Section~\ref{sec:moll:idealized}. We therefore construct the modified gradient
in two steps. The factor $(1-\eta)$ retains the raw pressure gradient away from
the shock and suppresses its sharp peak in the shock neighborhood. The removed
signed pressure change is not discarded: for each shock, it is reintroduced
as $\Delta p_{\ell}f_{\ell}$ and thereby spread over the prescribed finite
width. This gives
\begin{equation}
  \frac{\partial \tp}{\partial x}(x)
  \;=\;
  \underbrace{\bigl(1 - \eta(x)\bigr)\,\frac{\partial p}{\partial x}(x)}_{\substack{\text{smooth-background}\\\text{gradient}}}
  \;+\;
  \underbrace{\sum_{\ell}\Delta p_{\ell}\,f_{\ell}(x)}_{\substack{\text{shock-localized}\\\text{peak (PDF shaped)}}},
  \label{eq:mod_grad}
\end{equation}
The first term equals the raw pressure gradient wherever $\eta=0$ and fades
out as the shock is approached. The second term distributes the pressure
change $\Delta p_{\ell}$ according to the empirical PDF\@. By
Eq.~\eqref{eq:jump}, the suppressed gradient contribution integrates to
$\Delta p_{\ell}$ for each shock; because $f_{\ell}$ has unit area, the
replacement term reinserts exactly the same pressure change.
Eq.~\eqref{eq:mod_grad} is therefore the empirical counterpart of
Eq.~\eqref{eq:1d_p_tilde_grad}, augmented by the smooth background that is
absent from the idealized model.

\begin{figure}[!t]
  \centering
  \includegraphics[width=\textwidth]{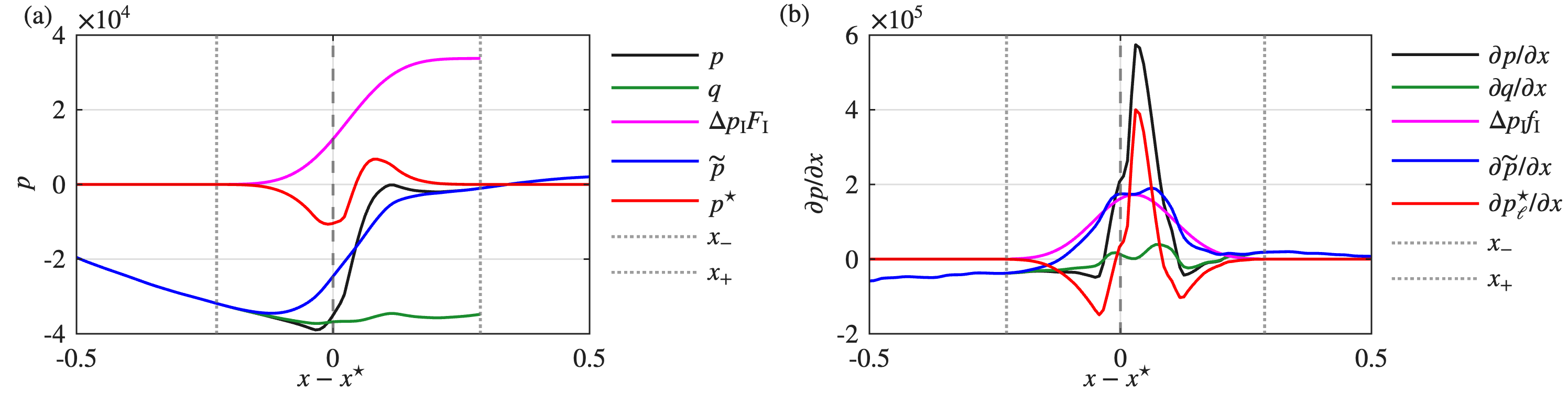}%
  \caption{Linewise components of the mollified--sharp decomposition for an
  upper-surface shock at $\alpha=2^{\circ}$ and $M_\infty=0.86$, extracted
  along the streamwise line $y=0.15$. Panel (a) shows the original pressure
  $p$, the smooth-background contribution $q$ from Eq.~\eqref{eq:p_tilde}, the
  accumulated shock contribution $\sum_{\ell}\Delta p_{\ell}F_{\ell}$, the
  mollified pressure $\tilde{p}$, and the local sharp correction $p^{\star}$ as
  functions of the relative coordinate $x-x^{\star}$. Panel (b) shows the
  corresponding gradients $\partial p/\partial x$, $\partial q/\partial x$,
  $\sum_{\ell}\Delta p_{\ell}f_{\ell}$, $\partial\tilde{p}/\partial x$, and
  $\partial p^{\star}_{\ell}/\partial x$. The dashed vertical line marks the
  detected shock location $x^{\star}$, and the dotted lines mark the limits
  $x_-$ and $x_+$ of the mollification interval.}
  \label{fig:mollified-sharp-line-components}
\end{figure}

Integrating Eq.~\eqref{eq:mod_grad} from a point $x_-$ upstream of
$\operatorname{supp}\eta$ gives the mollified pressure:
\begin{equation}
  \tp(x) \;=\; \underbrace{p(x_-)
  \;+\; \int_{x_-}^{x}\!\bigl(1 - \eta(x')\bigr)\,\frac{\partial p}{\partial x'}(x')\,dx'}_{\substack{\text{smooth-background}\\\text{contribution (}q(x) \text{ in Fig.~\ref{fig:mollified-sharp-line-components})}}}
  \;+\; \underbrace{\sum_{\ell}\Delta p_{\ell} \,F_{\ell}(x)}_{\substack{\text{mollified compression}\\\text{ramp (CDF shaped)}}}.
  \label{eq:p_tilde}
\end{equation}
The first bracketed term is the smooth-background contribution $q(x)$ depicted
in Fig.~\ref{fig:mollified-sharp-line-components}. The second term accumulates
each PDF-shaped gradient contribution into a CDF-shaped pressure ramp. Thus
Eq.~\eqref{eq:p_tilde} is the data-based counterpart of
Eq.~\eqref{eq:1d_p_tilde}: it preserves the smooth variation already present in
the CFD field while replacing each sharp transition by a finite-width ramp.

The detection of $\xstar_{\ell}(y)$ and its presence criteria are detailed in
Appendix~\ref{app:detection}, the construction of $\dstar_{\ell}$ and
$\fsig\ast\dstar_{\ell}$ in Appendix~\ref{app:graph-delta}, and the complete
grid-level realization in Appendix~\ref{app:algorithm}. The detector
calibration determines which lines contain a shock and where that shock is
placed, whereas $\sigma$ determines the width of its mollified representation.

\begin{figure}[!t]
  \centering
  \includegraphics[width=\textwidth]{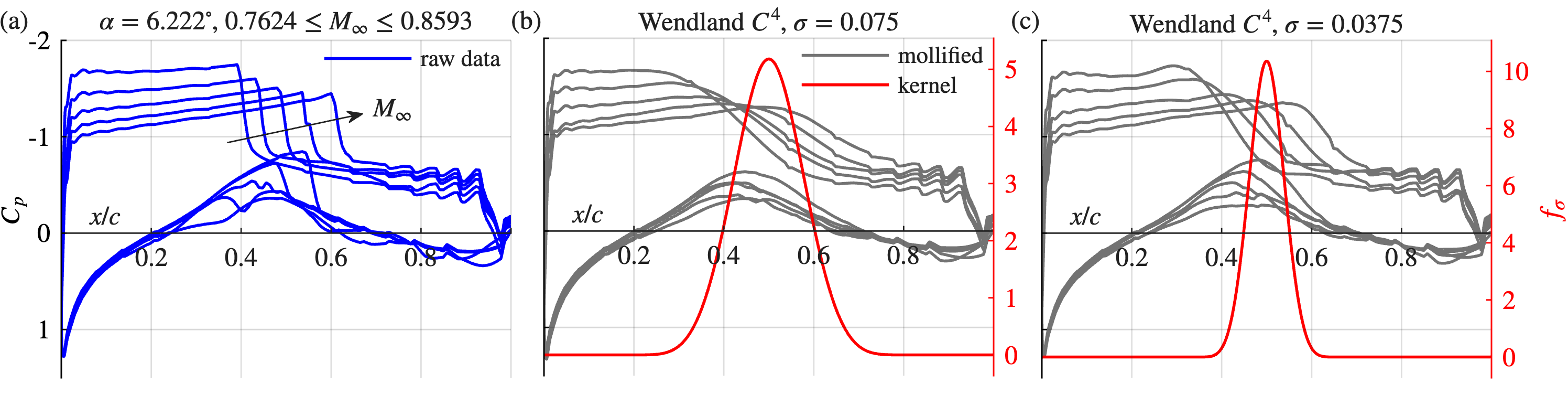}
  \caption{Illustration of the smoothing-width choice from the shock-location
  spacing across five neighboring training samples at
  $\alpha=6.222^{\circ}$ and increasing $M_\infty$. Panel (a) shows the raw
  surface pressure coefficient $C_p$. Panels (b) and (c) show the corresponding
  mollified profiles for the moment-matched Wendland $C^4$ kernel with
  $\sigma=0.075$ and $\sigma=0.0375$, respectively. The red reference kernels
  are centered at $x/c=0.5$ only to display their relative widths; the actual
  empirical density follows the detected shock location of each profile.
  Through Eq.~\eqref{eq:eta}, the same width determines the support and shape
  of the peak-normalized mollification weight $\eta_{\ell}$ in the
  isolated-shock limit.}
  \label{fig:cp-spacing-training}
\end{figure}

The width $\sigma$ is the main regularization scale. A smaller value produces a
tighter density and a sharper CDF ramp, leaving more of the moving sharp
feature in $\tp$ and slowing its POD decay. A larger value spreads the
transition over a wider region and generally accelerates the POD-energy decay,
at the cost of a more strongly mollified field.

The precise kernel shape is less consequential here than its width.
Preliminary comparisons for representative snapshots (not shown), using
moment-matched Gaussian and Wendland $C^4$ kernels, produced little discernible
difference in the mollified fields. A
Gaussian is nonzero at every distance, however, so a finite discrete
convolution requires an additional truncation radius. The Wendland kernel is
compactly supported by construction \citep{wendland1995piecewise}; it therefore
provides a finite support without an auxiliary cutoff. We use the Wendland
$C^4$ kernel for this practical reason.

For the RAE2822 data, $\sigma$ is chosen relative to the shock-location spacing
resolved by the retained training set. Fig.~\ref{fig:cp-spacing-training}
shows five neighboring samples along the Mach-number direction at fixed angle
of attack. The adopted value $\sigma=0.075$ gives the moment-matched Wendland
$C^4$ kernel a support radius of approximately $0.326$ chord lengths, which is
comparable to the representative shock displacement between neighboring
samples. Halving the width to $\sigma=0.0375$ reduces the support radius to
approximately $0.163$ chord lengths. For profiles with a pressure peak upstream
of the shock and a negative peak downstream, this narrower support retains
stronger local overshoots in the mollified profiles, as seen in
Fig.~\ref{fig:cp-spacing-training}(c). The wider value is therefore used for
the reported results.

Fig.~\ref{fig:mollified-sharp-line-components} complements the field-level
overview in Fig.~\ref{fig:decomposition-example} with a one-dimensional
breakdown along a streamwise line through a different upper-surface shock.
Panel (a) resolves the pressure-level quantities in
Eq.~\eqref{eq:p_tilde}: the smooth-background contribution $q$, the
CDF-shaped shock contribution $\sum_{\ell}\Delta p_{\ell}F_{\ell}$, and their
sum $\tilde{p}$, together with the original pressure $p$ and the local sharp
correction $p^{\star}$. Panel (b) shows the corresponding gradient-level
quantities in Eq.~\eqref{eq:mod_grad}. In particular, it displays the raw
pressure gradient, the retained background contribution, and the broader
PDF-shaped term that replaces the sharp shock peak. Thus the regularization is
applied to the pressure gradient through Eq.~\eqref{eq:mod_grad} and inherited
by the pressure field upon integration through Eq.~\eqref{eq:p_tilde}. The
remaining difference $p-\tilde{p}=p^{\star}$ is the local sharp correction
developed in Section~\ref{sec:sharp}.

The empirical construction retains the pressure gradient away from the
detected shocks and preserves the associated net pressure change while
replacing each localized shock gradient by a prescribed probability-density
profile. The residual required to recover the original sharp field is treated
next.

\FloatBarrier
\section{Local sharp correction}
\label{sec:sharp}

\begin{figure}[!t]
  \centering
  \includegraphics[width=\textwidth]{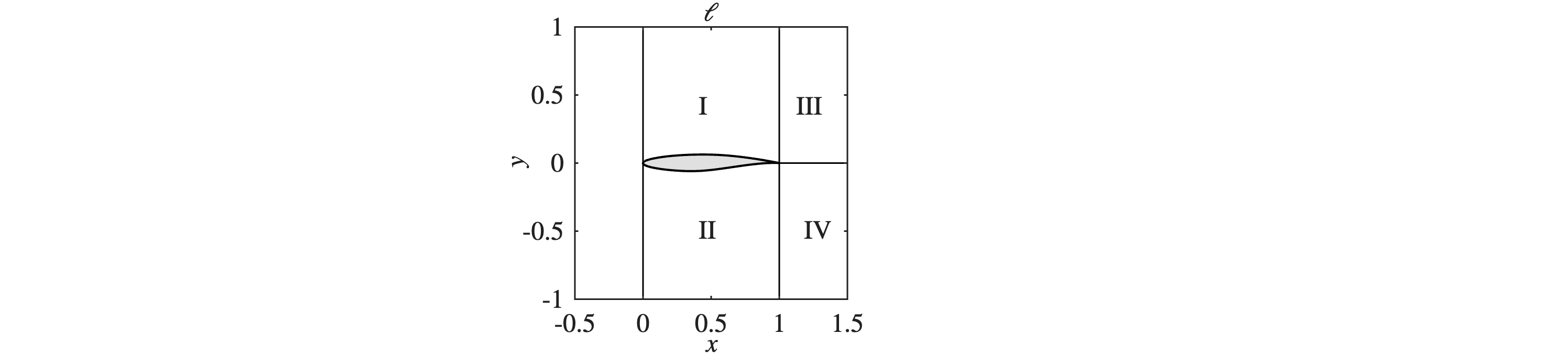}
  \caption{Fixed spatial partition used for the local sharp
  correction. Sectors $\sct{I}$ and $\sct{II}$ contain the upper and lower
  airfoil regions, respectively, for $0\leq x\leq1$, while sectors
  $\sct{III}$ and $\sct{IV}$ contain the corresponding downstream regions for
  $x>1$. The centerline separates the upper and lower sectors, and the
  upstream region $x<0$ is excluded from the shock model.}
  \label{fig:sector-definitions}
\end{figure}

A pressure snapshot may contain spatially separated sharp corrections
associated with shocks on the upper surface, lower surface, or in the wake. To
treat these moving features independently, we partition the physical domain
into the four fixed shock-bearing sectors shown in
Fig.~\ref{fig:sector-definitions}.

The index $\ell\in\{\sct{I},\sct{II},\sct{III},\sct{IV}\}$ identifies a sector
and, when a shock is detected there, its associated correction component. The
partition is chosen for the present database so that at most one detected
component is retained per sector. It is designed to keep independently moving
sharp features in separate local representations. The precise sector
boundaries are an implementation choice; other partitions can be used provided
that the individual components remain separated.

\subsection{Discrete component reconstruction}
\label{sec:sharp:component-reconstruction}

This sector partition splits the sharp correction $\pstar=p-\tp$ into local
components. For each component, the correction gradient is the difference
between the raw shock-gradient contribution removed during mollification and
its finite-width replacement:
\begin{equation}
  \frac{\partial \pstar_{\ell}}{\partial x}
  \;=\;
  \eta_{\ell}\,\frac{\partial p}{\partial x}
  \;-\; \Delta p_{\ell}\,f_{\ell},
  \qquad
  \pstar=\sum_{\ell}\pstar_{\ell}.
  \label{eq:pstar_components}
\end{equation}
After computing $\tp$, we first form the total residual
$\pstar=p-\tp$. Eq.~\eqref{eq:pstar_components} is then integrated along $x$
in each shock-containing sector and on each fixed-$y$ line, using a zero
upstream value. A final partition correction distributes the mismatch
introduced by discrete differentiation, integration, and endpoint correction
according to $\eta_{\ell}$. This step enforces
$\pstar=\sum_{\ell}\pstar_{\ell}$ on the Cartesian grid; its discrete form is
given in Appendix~\ref{app:algorithm}.

\subsection{Shock-aligned coordinates}
\label{sec:sharp:aligned-coordinates}

Each $\pstar_{\ell}$ is concentrated near a detected sharp feature whose
position, orientation, and spatial extent vary across the parameter space. We
therefore express each component in its own shock-aligned frame before
constructing the POD of Section~\ref{sec:surrogate}. The frame descriptors are
subsequently regressed over the parameter space together with the aligned
field coefficients. At a new parameter point, the predicted descriptors
define the inverse transformation that places, orients, and scales the
predicted sharp correction on the physical grid. The resulting POD spectra,
means, and leading modes in the aligned coordinates are shown later in
Fig.~\ref{fig:pod-mollified-secI}.

To construct the local frame, we use a spatial principal-axis analysis weighted
by the partitioned mollification weight $\eta_{\ell}$. Because
$\eta_{\ell}$ is derived from the peak-normalized shock-location probability
density, it retains the localized support and geometry needed for alignment.
Its weighted centroid determines the component position, the principal axes
determine its orientation, and the corresponding variances determine its
scales. Frames based on other localized shock fields were also tested, but
$\eta_{\ell}$ gave the most stable estimates in preliminary comparisons. The
alignment procedure itself is not tied to this choice.

On the sector-restricted support $\Omega_{\ell}$, the $\eta_{\ell}$-weighted
centroid and covariance are
\begin{equation}
  \xbar_{\ell}
  = \frac{\int_{\Omega_{\ell}}\xvec\,\eta_{\ell}(\xvec)\,d\xvec}
         {\int_{\Omega_{\ell}}\eta_{\ell}(\xvec)\,d\xvec},
  \qquad
  C_{\ell}
  = \frac{\int_{\Omega_{\ell}}(\xvec-\xbar_{\ell})(\xvec-\xbar_{\ell})^{\top}
           \eta_{\ell}(\xvec)\,d\xvec}
         {\int_{\Omega_{\ell}}\eta_{\ell}(\xvec)\,d\xvec},
  \label{eq:pca_cov}
\end{equation}
with eigenpairs ordered by variance,
\begin{equation}
  C_{\ell}\,\epar
  = \lambda_{\parallel,\ell}\,\epar,
  \qquad
  C_{\ell}\,\eperp
  = \lambda_{\perp,\ell}\,\eperp,
  \qquad
  \lambda_{\parallel,\ell}\geq\lambda_{\perp,\ell}.
  \label{eq:pca_axes}
\end{equation}
The eigenvector associated with the larger variance defines the dominant
alignment direction $\epar$, which approximates the mean shock-tangent
direction. The orthogonal eigenvector $\eperp$ defines the corresponding normal
direction. Their signs are chosen outward from the airfoil centerline and
downstream, respectively. The eigenvalues
$\lambda_{\parallel,\ell}$ and $\lambda_{\perp,\ell}$ are the
$\eta_{\ell}$-weighted positional variances and set the two spatial scales.
The dimensionless local coordinates, with multiplicative half-width factors
$\cperp$ and $\cpar$, are
\begin{equation}
  \xiperp(\xvec)
  =
  \frac{(\xvec-\xbar_{\ell})\cdot\eperp}
       {\cperp\sqrt{\lambda_{\perp,\ell \vphantom{\parallel}}}},
  \qquad
  \xipar(\xvec)
  =
  \frac{(\xvec-\xbar_{\ell})\cdot\epar}
       {\cpar\sqrt{\lambda_{\parallel,\ell}}}.
  \label{eq:xi}
\end{equation}

\begin{figure}[!t]
  \centering
  \includegraphics[width=\textwidth]{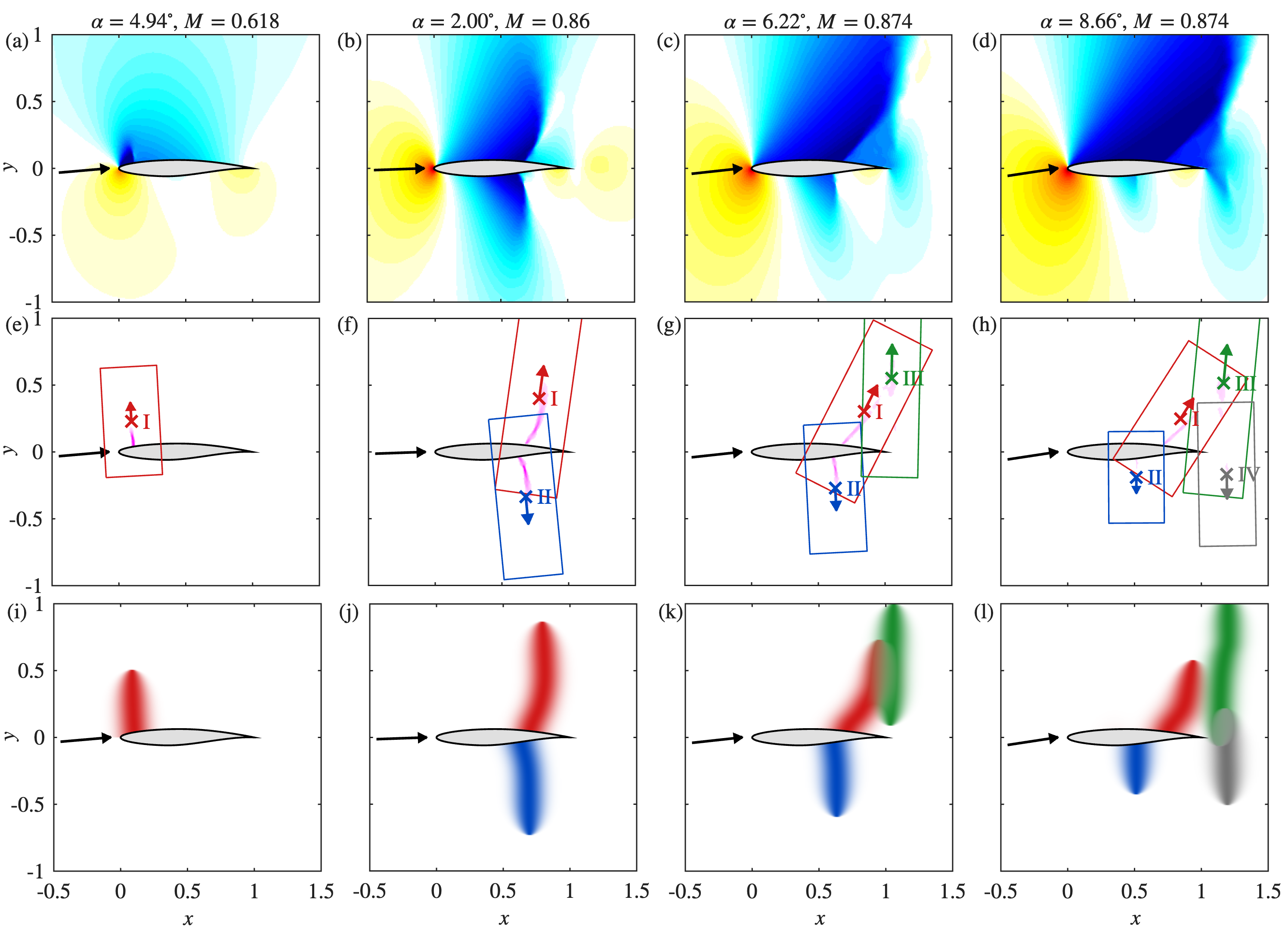}
  \caption{Representative sector decompositions and local alignment
  frames at four parameter points, where $\alpha$ is the angle of attack and
  $M_\infty$ the freestream Mach number: $\alpha=4.94^{\circ}$,
  $M_\infty=0.618$; $\alpha=2.00^{\circ}$, $M_\infty=0.86$;
  $\alpha=6.22^{\circ}$, $M_\infty=0.874$; and
  $\alpha=8.66^{\circ}$, $M_\infty=0.874$, with one, two, four, and four
  sectors with detected shocks, respectively. The upper row shows the
  pressure fields $p(x,y)$, with color limits adjusted for visual clarity. The
  middle row shows the sector-restricted shock indicators and traces together
  with the resulting local frames: the centroids $\xbar_{\ell}$, tangent and
  normal axes, and the physical footprints of the local coordinate domains.
  The frames are overlaid on the indicators for legibility; in the reported
  implementation, they are computed from the partitioned mollification
  weights $\eta_{\ell}$ shown in the lower row through
  Eqs.~\eqref{eq:pca_cov}--\eqref{eq:xi}. The airfoil outline and
  freestream-direction arrow are included for orientation.}
  \label{fig:sector-eta-components}
\end{figure}

Fig.~\ref{fig:sector-eta-components} spans representative shock configurations
in the retained database. The first example contains a single
upper-surface shock component, and the second contains components on both the
upper and lower surfaces. The final two examples show more complex
multi-component trailing-edge shock systems that extend into all four
airfoil and downstream sectors. Together, the examples illustrate the
progression from one shock-bearing sector to the observed maximum of four shown in
Fig.~\ref{fig:parameter-space}. Three-sector cases use the same construction
with one sector absent, while a sample with no detected component requires no
local frame. Because the fixed partition may divide a connected shock system
across a sector boundary, the number of sectors with detected shocks does not necessarily
equal the number of physically distinct shock waves.

The figure also illustrates the physical interpretation of the transform.
The $\eta_{\ell}$-weighted centroid $\xbar_{\ell}$ maps to the local origin
$\xivec=\bm{0}$, about which the sharp-correction component is represented. It
is not, in general, the centroid of the sign-changing field $\pstar_{\ell}$
itself. The dominant principal axis of $\eta_{\ell}$ supplies the direction
$\epar$ used for alignment; the orthogonal axis $\eperp$ resolves
variation across the detected feature. Each snapshot's
component $\pstar_{\ell}$ is then interpolated onto the fixed grid in $\xivec$
through its own transform, Eq.~\eqref{eq:xi}. The component location,
orientation, and scale are thereby carried by the transform rather than the
aligned field. The sector-dependent axis conventions, scale factors, and
local-grid interpolation used in the reported implementation are specified in
Appendix~\ref{app:algorithm}.

\FloatBarrier
\section{Database and parametric reduced-order model}
\label{sec:surrogate}

Sections~\ref{sec:moll} and~\ref{sec:sharp} define the decomposition and
alignment for an individual pressure snapshot. We apply that construction to
a database spanning $\muvec=(\alpha,M_\infty)$ and approximate the parameter
dependence of the resulting fields and alignment quantities. The database
selection is described first,
followed by the POD representations and the regressions used for evaluation at
an unsampled parameter point.

\subsection{Database and retained subset}
\label{sec:database}

The results use the transonic RAE2822 airfoil database of
\citet{catalani2023comparative}. The database contains more than 2000
pressure fields from two-dimensional RANS solutions around the RAE2822 airfoil
at Reynolds number $Re=6.5\times10^{6}$, spanning freestream Mach numbers up
to $M_\infty=0.9$ and angles of attack up to $9^{\circ}$. The original database
provides separate parameter sets generated by Clenshaw--Curtis sampling for
training and by random sampling for testing. These sets retain their original
roles here.

The present study retains the full available angle-of-attack range but
restricts both sets to $M_\infty\geq0.615$. With the indicator used here, no
shocks are detected below this threshold. For such samples the indicator vanishes, the
mollified--sharp decomposition reduces to $\tp=p$ and $\pstar=0$, and the
additional shock treatment is inactive. Including these predominantly smooth
cases would therefore add little to the present assessment of the proposed
method. The cutoff is specific to this database and purpose; it is not a
restriction of the decomposition. Samples with nonfinite pressure or flow
quantities required by the shock indicator are also removed. This leaves 332
training and 327 testing samples, as summarized in
Table~\ref{tab:retained-database-subset}. All retained fields are subsequently
interpolated to the common Cartesian grid introduced in
Section~\ref{sec:moll:empirical}.

\begin{figure}[!t]
  \centering
  \includegraphics[width=\textwidth]{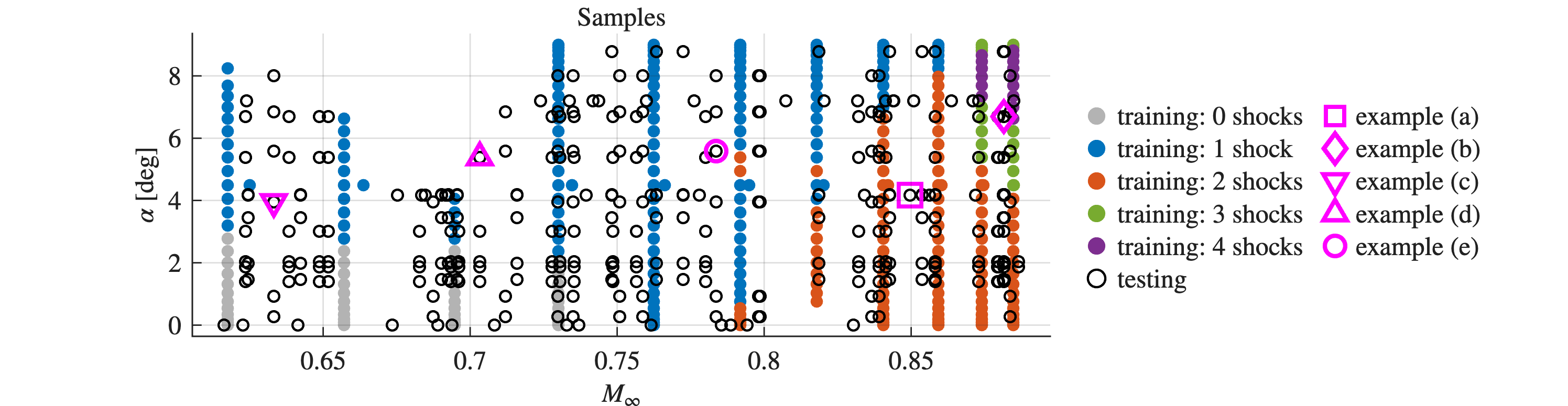}
  \caption{Parameter-space distribution of the retained RAE2822
  samples. Filled symbols show the training samples, colored by the number of
  detected shock-bearing sectors, and open circles show the testing samples.
  The magenta symbols identify the five selected testing cases examined
  in Fig.~\ref{fig:model-cp-p-five-row-test}.}
  \label{fig:parameter-space}
\end{figure}

The retained parameter distribution is shown in
Fig.~\ref{fig:parameter-space}. The color of each training point gives the
number of sectors in which a shock is detected. Shock-free
samples remain present above the Mach-number cutoff, particularly near the
onset of transonic flow, and provide the transition between the smooth and
shock-bearing regimes. The testing points are used only for evaluation.

\begin{center}
\begin{minipage}{\textwidth}
\centering
\renewcommand{\arraystretch}{1.2}
\begin{tabular}{@{}lccc@{}}
\toprule
\textbf{Subset} & \textbf{Sampling} & \textbf{Available samples} & \textbf{Retained samples} \\
\midrule
Training & Clenshaw--Curtis & 1012 & 332 \\
Testing  & random           & 1007 & 327 \\
\bottomrule
\end{tabular}
\par\medskip
\captionof{table}{Retained transonic subset of the RAE2822 pressure-field
database. The selection retains the available angle-of-attack range subject
to $M_\infty\geq0.615$ and finite pressure and flow data.}
\label{tab:retained-database-subset}
\end{minipage}
\end{center}

The shock indicator is evaluated for all 332 retained training samples. Of
these, 284 contain at least one detected shock-bearing sector. A single
parameter point may contain detected shocks in several sectors, so the counts in
Table~\ref{tab:sector-snapshot-counts} are occurrence counts and do not sum to
the number of shocked parameter points.
\begin{center}
\begin{minipage}{\textwidth}
\centering
\renewcommand{\arraystretch}{1.2}
\begin{tabular}{@{}lcc@{}}
\toprule
\textbf{Sector} & \textbf{Shocked training samples} & \textbf{Aligned sharp snapshots} \\
\midrule
Upper airfoil     & 284 & 284 \\
Lower airfoil     & 128 & 126 \\
Downstream upper  &  33 &  33 \\
Downstream lower  &  23 &  18 \\
\bottomrule
\end{tabular}
\par\medskip
\captionof{table}{Sector-wise shock content of the retained training subset.
The first count requires a positive shock detection in the sector. The second
additionally requires a valid alignment transform and nonempty coverage of
the mapped alignment weight $\eta_{\ell}$ on the local grid.}
\label{tab:sector-snapshot-counts}
\end{minipage}
\end{center}

The two columns of Table~\ref{tab:sector-snapshot-counts} reflect different
requirements. Identifying a shock in a sector only requires that the indicator
satisfy the detection criteria in Appendix~\ref{app:detection}. Constructing
an aligned sharp snapshot additionally requires a nonzero
$\eta_{\ell}$ field, finite centroid and principal-axis quantities in
Eqs.~\eqref{eq:pca_cov} and~\eqref{eq:pca_axes}, and nonempty coverage after
mapping $\eta_{\ell}$ to the fixed local grid using Eq.~\eqref{eq:xi}. Two
lower-airfoil and
five downstream-lower detections do not satisfy all of these alignment
requirements and are omitted from the corresponding local POD ensembles
rather than assigned an unreliable frame. Their original binary indicator values
are nevertheless retained when fitting the shock-presence function, which
uses all 332 training parameter points.
\FloatBarrier

\subsection{POD--GPR construction}
\label{sec:surrogate:pod-gpr}

For each retained training parameter point
$\muvec_k=(\alpha_k,M_{\infty,k})$, the empirical construction of
Eqs.~\eqref{eq:fl}--\eqref{eq:p_tilde} produces the global mollified pressure
$\tp_k$ on the Cartesian grid. The residual $\pstar_k=p_k-\tp_k$ is partitioned
according to Eq.~\eqref{eq:pstar_components}. Each component with a valid frame
is then mapped through Eq.~\eqref{eq:xi} to the same local
$(\xi_{\perp},\xi_{\parallel})$ grid. This procedure gives one global snapshot
ensemble for $\tp$ and four sector-specific ensembles for $\pstar_{\ell}$. The
global ensemble contains every retained training sample; each local ensemble
contains only the aligned snapshots listed in
Table~\ref{tab:sector-snapshot-counts}.

POD is applied independently to these five ensembles. After subtracting the
appropriate snapshot mean, a singular-value decomposition provides the basis,
and the smallest rank that captures 99.9\% of the snapshot energy is retained.
The projections onto these bases give modal coefficients at the sampled
values of $\muvec$. A separate Gaussian-process regression (GPR) model is then
fitted to each coefficient as a function of $\alpha$ and $M_\infty$. Thus, for
the mollified part, with mean $\bar{\tp}$ and rank-$r$ basis
$\{\phi_j\}_{j=1}^{r}$,
\begin{equation}
  \tpr(\xvec; \muvec)
  \;=\; \bar{\tp}(\xvec) \;+\; \sum_{j=1}^{r} \hat{a}_{j}(\muvec)\,\phi_j(\xvec).
  \label{eq:p_tilde_hat}
\end{equation}
For each aligned component, with mean $\bar{p}^{\star}_{\ell}$ and
rank-$r_{\ell}$ basis
$\{\psi_{\ell,j}\}_{j=1}^{r_{\ell}}$,
\begin{equation}
  \pstarr_{\ell}(\xivec;\muvec)
  =
  \bar{p}^{\star}_{\ell}(\xivec)
  + \sum_{j=1}^{r_{\ell}} \hat{b}_{\ell,j}(\muvec)\,\psi_{\ell,j}(\xivec).
  \label{eq:p_star_hat_aligned}
\end{equation}
Here, $\hat{a}_{j}$ and $\hat{b}_{\ell,j}$ denote the regressed coefficients
at the requested parameter point. Eq.~\eqref{eq:p_tilde_hat} is evaluated
directly on the physical grid. Eq.~\eqref{eq:p_star_hat_aligned}, by contrast,
first reconstructs the shape and amplitude of a sharp-correction component in
its local frame; its position, orientation, and scale are supplied separately
by the alignment descriptors defined in Eq.~\eqref{eq:theta_alignment} below.

\begin{figure}[!t]
  \centering
  \includegraphics[width=\textwidth]{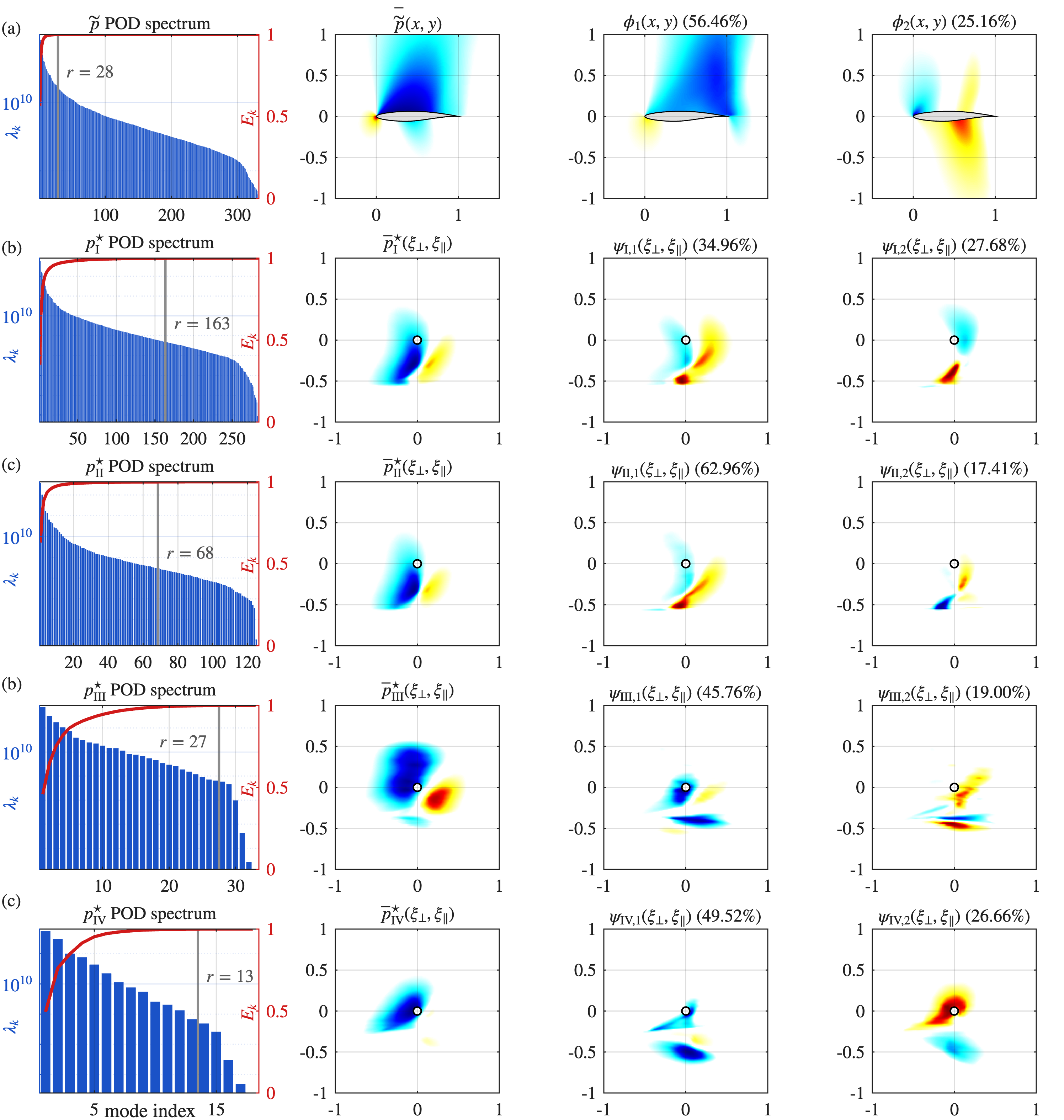}
  \caption{POD spectra, means, and leading basis functions for the global
  mollified pressure and the four aligned sharp-correction sectors. The rows
  correspond to $\tp$, $\pstar_{\sct{I}}$, $\pstar_{\sct{II}}$,
  $\pstar_{\sct{III}}$, and $\pstar_{\sct{IV}}$. In each row, the first
  column shows the POD eigenvalues $\lambda_k$ (blue bars), cumulative energy
  $E_k$ (red curve), and the 99.9\% energy-truncation rank (gray line). The
  remaining columns show the snapshot mean and the first two modes. The
  mollified-pressure row is plotted in the physical coordinates $(x,y)$ as
  $\bar{\tp}$, $\phi_1$, and $\phi_2$; the sector rows are plotted in the
  aligned coordinates $(\xi_{\perp},\xi_{\parallel})$ as
  $\bar{p}^{\star}_{\ell}$, $\psi_{\ell,1}$, and $\psi_{\ell,2}$. Percentages
  in the mode titles give the individual modal energy fractions.}
  \label{fig:pod-mollified-secI}
\end{figure}

Fig.~\ref{fig:pod-mollified-secI} summarizes the five PODs in the coordinate
systems in which they are constructed. At the 99.9\% cumulative POD-energy
threshold, the mollified-pressure ensemble requires $r=28$ modes, while
sectors $\sct{I}$--$\sct{IV}$ require $r_{\ell}=163$, 68, 27, and 13 modes,
respectively. These ranks determine the numbers of GPR models for the modal
coefficients appearing in
Eqs.~\eqref{eq:p_tilde_hat} and~\eqref{eq:p_star_hat_aligned}. The means and
leading modes also show the distinction between the global representation of
$\tp$ and the local, aligned representations of the sharp corrections.

The aligned field coefficients alone do not determine where a component must
be placed in the physical domain. For this purpose, the geometric quantities
defined in Section~\ref{sec:sharp:aligned-coordinates} are collected in the
parameter-dependent frame descriptor
\begin{equation}
  \theta_{\ell}(\muvec)
  =
  \bigl(
    \bar{x}_{\ell},\,\bar{y}_{\ell},\,
    \log\lambda_{\parallel,\ell},\,\log\lambda_{\perp,\ell},\,
    e_{\parallel,x,\ell},\,e_{\parallel,y,\ell}
  \bigr),
  \label{eq:theta_alignment}
\end{equation}
where $\epar=(e_{\parallel,x,\ell},e_{\parallel,y,\ell})$.
Here, $\muvec=(\alpha,M_\infty)$ is the two-dimensional parameter vector, and
$\theta_{\ell}$ is a six-component descriptor of the local frame. When a shock
is predicted in sector $\ell$, the centroid
$(\bar{x}_{\ell},\bar{y}_{\ell})$ places its sharp-correction component, the
direction $\epar$ orients it, and the square roots of the eigenvalues set its
local length scales.
These quantities define the inverse of Eq.~\eqref{eq:xi}, which maps the
regressed aligned component back to the physical grid. Adding this component
through Eqs.~\eqref{eq:p_star_hat_combined} and~\eqref{eq:p_hat} then restores
the localized sharp structure to the predicted mollified pressure. The
logarithms of the two eigenvalues are regressed so that
exponentiation of their predictions yields positive eigenvalues and,
therefore, positive scale factors in Eq.~\eqref{eq:xi}.
Eq.~\eqref{eq:theta_alignment} lists the physical frame quantities. For
regression, its final two entries are replaced by the sign-invariant pair
$(\cos 2\vartheta_{\ell},\sin 2\vartheta_{\ell})$, with
$\epar=(\cos\vartheta_{\ell},\sin\vartheta_{\ell})$. This doubled-angle form
removes the directional ambiguity of a principal axis. After prediction,
$\vartheta_{\ell}=\tfrac{1}{2}\operatorname{atan2}(\sin2\vartheta_{\ell},\cos2\vartheta_{\ell})$,
and the signs are selected consistently with the sector geometry before
Eq.~\eqref{eq:xi} is inverted to return
$\pstarr_{\ell}$ to the physical grid.

A separate GPR model $\hat{\chi}_{\ell}(\muvec)$ determines whether sector
$\ell$ contributes at the requested parameter point. It is fitted to the
binary indicator values $\chi_{\ell,k}\in\{0,1\}$ defined in
Appendix~\ref{app:detection}, including both shocked and shock-free training
samples. A component is included when
$\hat{\chi}_{\ell}(\muvec)\geq0.5$. The continuous GPR output is clipped to
$[0,1]$, but is used only for this threshold comparison and is not interpreted
as a calibrated probability. The complete construction and evaluation
procedure is given in Appendix~\ref{app:algorithm}.
When included, the component's modal coefficients and frame descriptor are
evaluated, the aligned field is reconstructed by
Eq.~\eqref{eq:p_star_hat_aligned}, and the inverse alignment transform places
it on the physical grid. A component is discarded if its predicted centroid
lies outside the corresponding sector. The physical-grid sector components
then add directly,
\begin{equation}
  \pstarr(\xvec; \muvec)
  \;=\;
  \sum_{\ell}
  \pstarr_{\ell}(\xvec; \muvec),
  \label{eq:p_star_hat_combined}
\end{equation}
where $\ell$ runs over the four sectors. No additional blending is required:
the component construction in Eq.~\eqref{eq:pstar_components} forms an exact
partition of the training correction, consistent with
$\sum_{\ell}\eta_{\ell}=\eta$. Finally, the predicted pressure is assembled as
\begin{equation}
  \pr(\xvec; \muvec)
  \;=\;
  \tpr(\xvec; \muvec) + \pstarr(\xvec; \muvec).
  \label{eq:p_hat}
\end{equation}
Evaluation at an unsampled parameter point therefore requires four groups of
regressed quantities: the mollified-pressure coefficients
$\hat{a}_j$, the aligned sharp-correction coefficients
$\hat{b}_{\ell,j}$, the entries of $\theta_{\ell}$, and the sector-presence
functions $\hat{\chi}_{\ell}$. One GPR is fitted for each quantity, as listed
in Table~\ref{tab:reg}. GPR is used here as a convenient smooth regression
method; the decomposition and reconstruction do not depend on this particular
choice, and another suitable regression method could be substituted. Common
alternatives include radial-basis-function interpolation and polynomial chaos
expansions.

\begin{center}
\begin{minipage}{\textwidth}
\centering
\renewcommand{\arraystretch}{1.25}
\begin{tabular}{@{}p{0.10\textwidth}p{0.25\textwidth}p{0.52\textwidth}@{}}
\toprule
\textbf{Quantity} & \textbf{Number of GPRs} & \textbf{Role} \\
\midrule
$\hat{a}_{j}(\muvec)$
  & $r=28$ & global mollified-pressure modal coefficients, $j=1,\ldots,r$ \\
$\hat{b}_{\ell,j}(\muvec)$
  & $\sum_{\ell} r_{\ell}=271$ & aligned sharp-correction modal coefficients, $j=1,\ldots,r_{\ell}$ \\
$\theta_{\ell}(\muvec)$
  & $4\times 6=24$ & alignment-transform descriptors of Eq.~\eqref{eq:theta_alignment} \\
$\hat{\chi}_{\ell}(\muvec)$
  & $4\times 1=4$ & shock-presence GPRs fitted to the indicators $\chi_{\ell,k}$ of Eq.~\eqref{eq:shock_presence_indicator} \\
\bottomrule
\end{tabular}
\par\medskip
\captionof{table}{GPR models for the parameter-dependent quantities of the
global mollified field and the four sector-local sharp corrections. The POD
counts correspond to the 99.9\% energy-truncation ranks shown in
Fig.~\ref{fig:pod-mollified-secI}.}
\label{tab:reg}
\end{minipage}
\end{center}
\FloatBarrier

\FloatBarrier
\section{Results}
\label{sec:results}

The preceding sections illustrated the mollified--sharp decomposition on
individual pressure snapshots, introduced the shock-aligned local frames, and
constructed the resulting POD representations. We now assess the complete
online reconstruction of Eq.~\eqref{eq:p_hat}. We first examine five selected
testing conditions that were not used to construct the model. They include
representative and deliberately challenging flow regimes. We then
compare the method over the full retained training and testing sets with a
standard POD--GPR model constructed directly from the undecomposed pressure
fields.

\subsection{Selected testing cases}
\label{sec:results:representative-cases}

\begin{figure}[p]
  \centering
  \includegraphics[width=\textwidth]{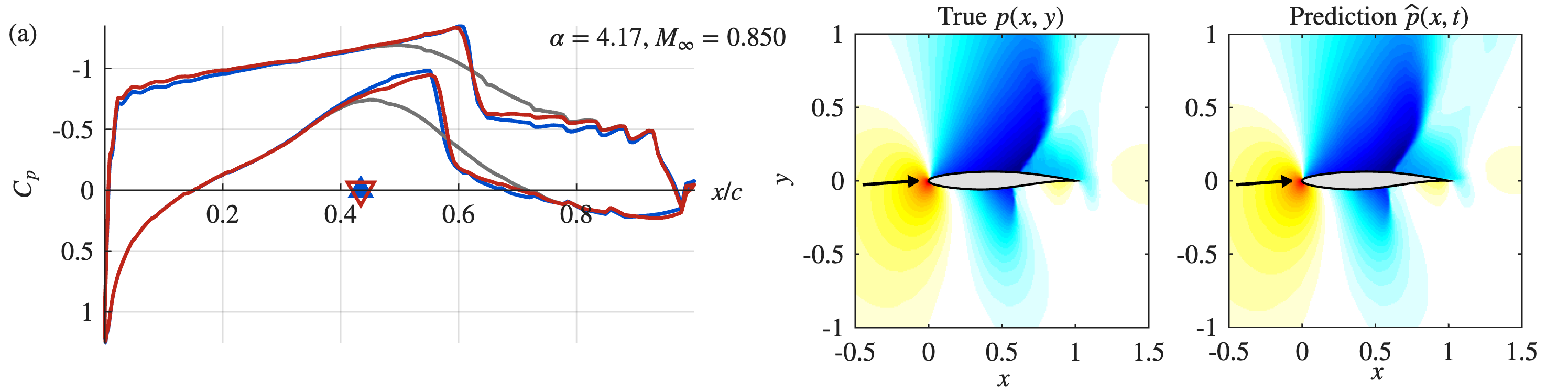}\\[-1mm]
  \includegraphics[width=\textwidth]{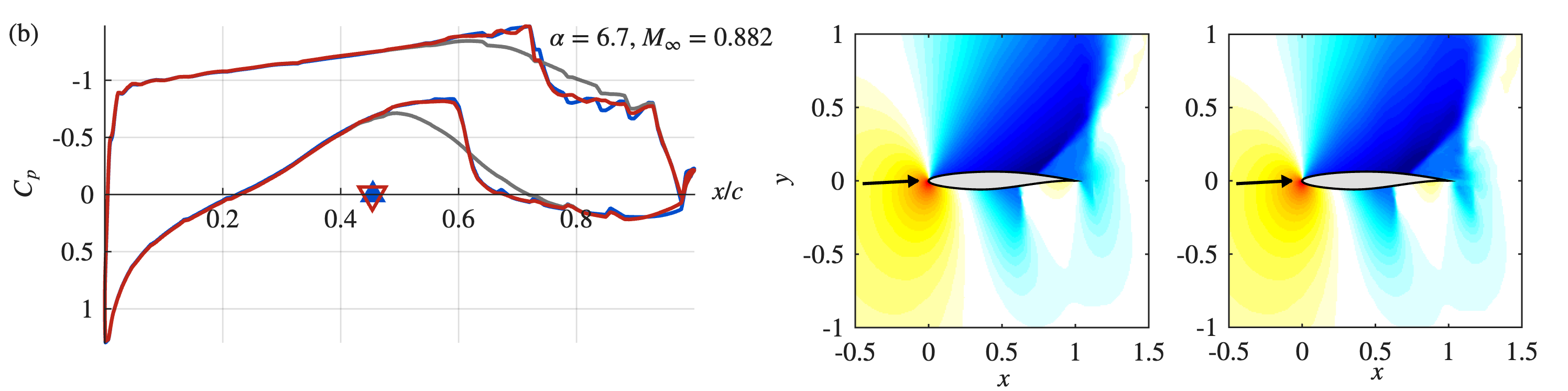}\\[-1mm]
  \includegraphics[width=\textwidth]{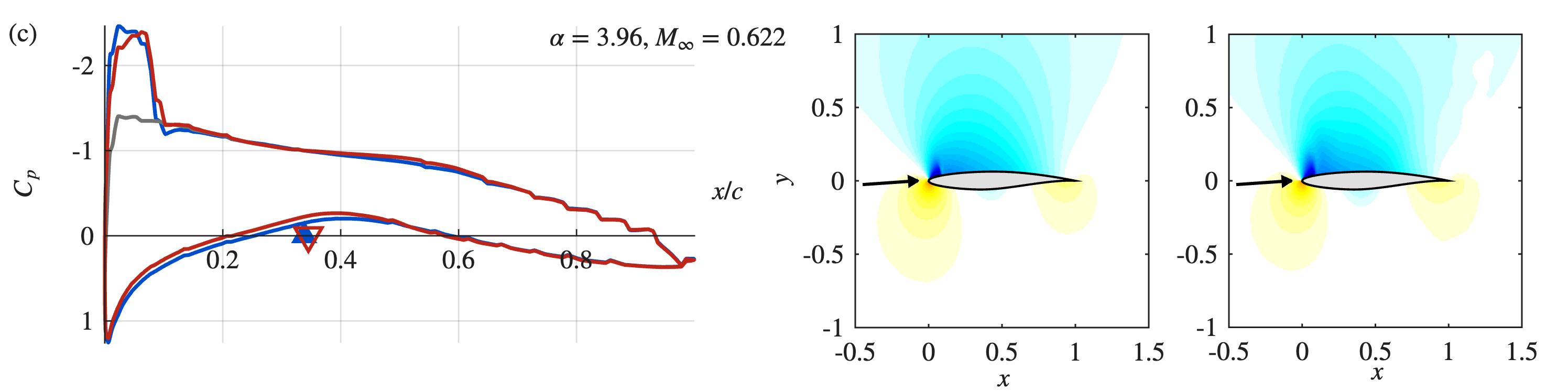}\\[-1mm]
  \includegraphics[width=\textwidth]{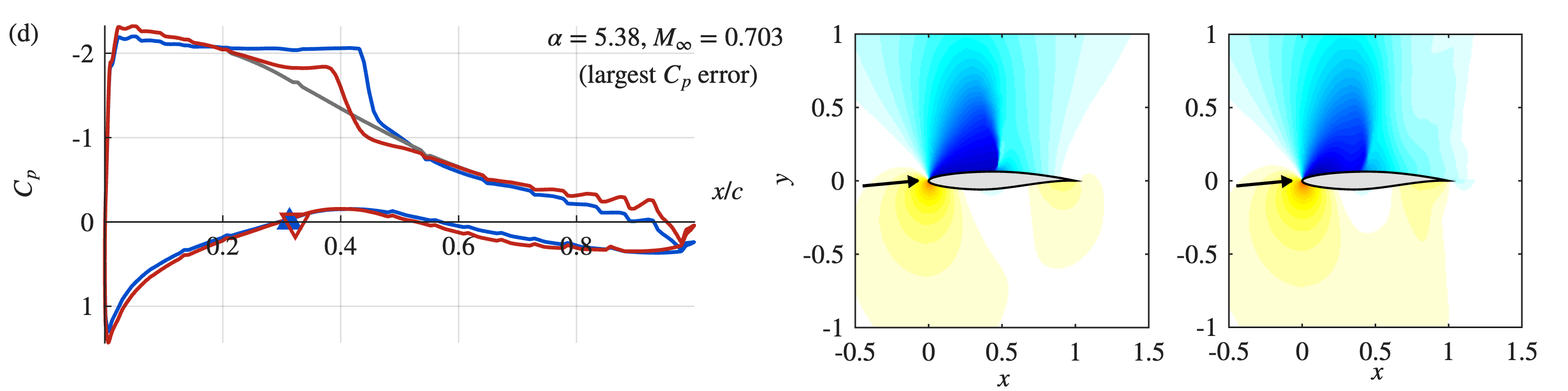}\\[-1mm]
  \includegraphics[width=\textwidth]{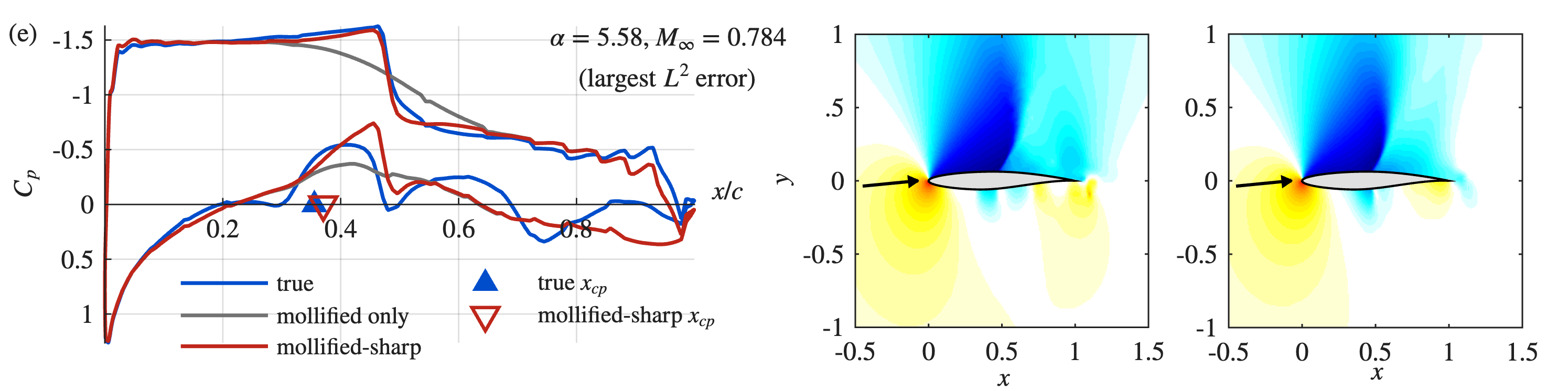}
  \par\vspace{-2mm}
  \caption{Selected test-set predictions of the mollified--sharp POD--GPR
  model. The columns show surface $C_p$ (left; reference in blue,
  mollified-only in gray, and final prediction in red), reference pressure
  (center), and predicted pressure (right); triangles mark the extracted shock
  coordinates. Rows (a) and (b) contain representative multi-shock cases, row
  (c) a small leading-edge shock, and rows (d) and (e) the largest
  surface-$C_p$ and full-field relative-$L^2$ testing errors. Standard
  POD--GPR predictions for the same cases are shown in
  Fig.~\ref{fig:model-cp-p-stdpod-five-row}.}
  \label{fig:model-cp-p-five-row-test}
\end{figure}

The selected parameter points are marked by magenta symbols in
Fig.~\ref{fig:parameter-space}. Unlike the snapshots used earlier to explain
the decomposition and alignment, these are predictions at retained testing
conditions. The first three were chosen to illustrate distinct shock
configurations, whereas the final two are the cases with the largest
surface-pressure and full-field testing errors, respectively.
Fig.~\ref{fig:model-cp-p-five-row-test} therefore combines representative
cases with two worst-error cases rather than
presenting all five as typical examples.

Rows (a)--(c) show the intended effect of the reconstruction. The gray
mollified-only profiles retain the slowly varying pressure distribution but
spread the shock-induced changes over a finite streamwise interval. Adding the
predicted local correction restores the sharp transitions and brings the red
profiles close to the references. This remains true for the multiple
shock configurations in rows (a) and (b), which contain upper-surface,
lower-surface, and downstream shocks, and for the small leading-edge shock near
the onset of the retained transonic regime in row (c). The corresponding
full-field plots show that the reduced-order model also preserves the principal
pressure structures away from the airfoil surface.

Row (d), marked by the upward triangle in
Fig.~\ref{fig:parameter-space}, occurs at
$(M_\infty,\alpha)=(0.703,5.38^\circ)$. The predicted upper-surface shock is
slightly upstream of the reference shock and its pressure change is too weak,
which explains much of the largest testing-set surface-$C_p$ error. This point
lies in a sparsely populated neighborhood immediately on the high-Mach side of
zone I identified by \citet{catalani2023comparative}. Those authors describe
zone I as a high-loading, unsteady regime in which the shock position is
highly sensitive to perturbations; RANS cases that did not converge in that
zone were excluded from their training and validation databases. Although the
present testing solution is available, its proximity to that excluded regime
makes the otherwise close reconstruction noteworthy.

Row (e), marked by the circle in Fig.~\ref{fig:parameter-space}, occurs at
$(M_\infty,\alpha)=(0.784,5.58^\circ)$. Here the primary upper-surface shock
is reproduced reasonably well, but the reference $C_p$ contains pronounced
oscillations over the downstream lower surface and weaker irregularities on
the upper surface that are not reproduced by the model. Inspection of the
pressure field suggests that these fluctuations are not a stable,
well-resolved mean-flow feature. This interpretation is consistent with
\citet{catalani2023comparative}, who specifically note that cases at
$\alpha=5.58^\circ$ and $M_\infty\in[0.7,0.8]$ lie close to their unsteady CFD
cases and may exhibit peculiar flow fields. The oscillations may therefore
reflect residual unsteadiness or incomplete convergence of a nominally
steady RANS solution. We nevertheless retain the snapshot as supplied and
include its error without modification. Row (e) therefore represents both a
limiting prediction case and a case for which the fidelity of the nominally
steady reference solution is less certain.

\subsection{Quantitative comparison with standard POD--GPR}
\label{sec:results:standard-pod-comparison}

The baseline is a standard POD--GPR model constructed directly from the
original pressure snapshots, without mollification, shock-aligned local
corrections, or the additional shock-presence and frame regressions. It uses the same 332
training and 327 testing samples, the same Cartesian grid, the same 99.9\%
cumulative POD-energy criterion, and the same GPR procedure as the proposed
model. This criterion retains 59 modes for the standard pressure basis, so 59
modal-coefficient GPRs are fitted and evaluated. Increasing the retained rank
in additional standard-POD calculations did not materially reduce the
testing errors; the rank study and computational-cost comparison are reported
in Appendix~\ref{app:cost}. Thus, the comparison retains the data and GPR
procedure while changing the representation used for shock-bearing fields.

Two complementary types of error are considered. The full-field relative
$L^2$ error is the discrete norm
$\|\hat{p}-p\|_2/\|p\|_2$ over finite Cartesian-grid nodes outside the airfoil.
Because the grid is uniform, the constant cell-area factor cancels from this
ratio. The database provides kinematic gauge pressure, and the pressure
coefficient is evaluated as
$C_p=p/[0.5(M_\infty a_\infty)^2]$, with $a_\infty=340~\mathrm{m\,s^{-1}}$.
The airfoil contour is resampled uniformly in arc length using
$\max(400,2N_{\mathrm{wall}})$ points, where $N_{\mathrm{wall}}$ is the
number of source wall points. The Cartesian pressure fields are interpolated
linearly to this contour after nearest-exterior filling inside the airfoil.
The surface metric is the arithmetic mean of
$|\hat{C}_p-C_p|$ over these points. These aggregate measures are supplemented
by shock-specific quantities because a field norm alone does not distinguish
shock-placement error from smooth-background error.

\begin{figure}[!t]
  \centering
  \includegraphics[width=\textwidth]{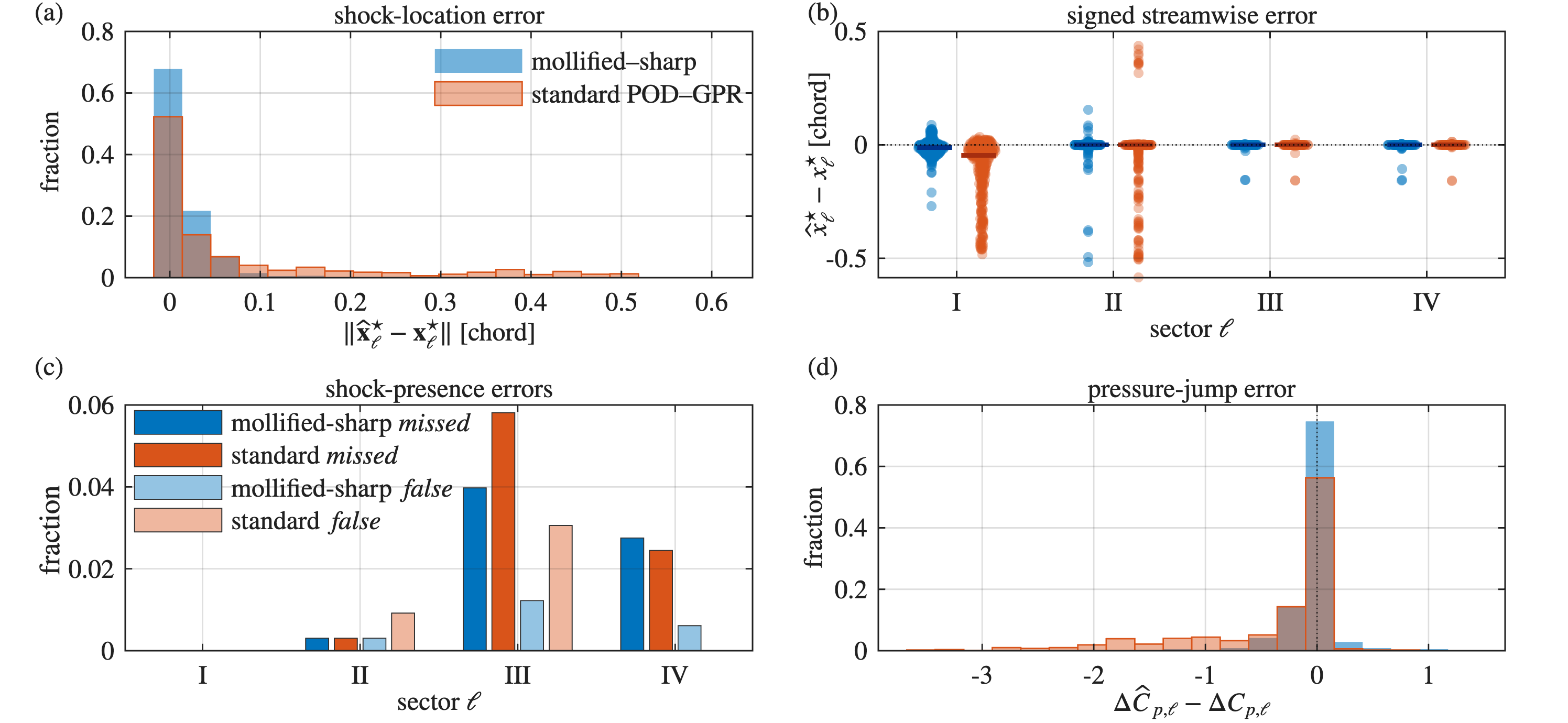}
  \caption{Shock-feature comparison between the mollified--sharp
  POD--GPR model and the standard POD--GPR baseline on the testing set. Panel
  (a) shows the Euclidean shock-centroid error distribution for sector--sample
  pairs in which the reference and prediction both contain a detected shock.
  Panel (b) shows the corresponding signed streamwise centroid error
  $\hat{x}^{\star}_{\ell}-x^{\star}_{\ell}$ by sector; markers are plotted as
  swarm plots and the horizontal line segments indicate sector-wise medians.
  Panel (c) reports missed and false shock-presence detections as percentages
  of testing cases in each sector. Panel (d) shows the error in the
  postprocessed pressure-jump estimate $\Delta C_{p,\ell}$.}
  \label{fig:shock-feature-comparison}
\end{figure}

For Fig.~\ref{fig:shock-feature-comparison}, the same pressure-only
postprocessing is applied to every reference and predicted testing field. The
magnitude of the Cartesian pressure gradient is normalized by its fieldwise
maximum. Within each sector, values above 0.125 weight the shock centroid, and
a shock is recorded when the sector maximum reaches 0.125. The pressure change
is evaluated on each contributing fixed-$y$ line between points 0.0375 chord
lengths upstream and downstream of its weighted streamwise center, then
averaged using the line weights. Panels (a), (b), and (d) include only
sector--sample pairs for which both fields contain a detected shock. The missed
and false detections in panel (c) are reported as percentages of all testing
cases in the corresponding sector.

The location errors of the mollified--sharp model in
Fig.~\ref{fig:shock-feature-comparison}(a) are concentrated close to zero,
whereas the standard model exhibits a substantially broader tail. The
sector-wise view in panel (b) shows that the largest standard-model errors
occur primarily in the airfoil sectors $\sct{I}$ and $\sct{II}$; the proposed
model reduces both their typical magnitude and the number of large
misplacements, although isolated outliers remain. The most prominent of these
are associated with the challenging data regimes discussed in connection with
Fig.~\ref{fig:model-cp-p-five-row-test}. The lower-surface feature in row (e)
produces the largest location error, and the row (d) case produces another
large error. Further outliers occur among nearby $\alpha=5.58^\circ$ cases and
near the high-Mach failure zone II identified by
\citet{catalani2023comparative}. Panel (c) shows that missed and false
shock detections remain below approximately 6\% in every sector for both
methods. The mollified--sharp model produces fewer false detections in sectors
$\sct{II}$ and $\sct{III}$ and fewer missed detections in sector
$\sct{III}$. The two methods agree in sector $\sct{I}$, whereas the standard
model performs slightly better in sector $\sct{IV}$. Finally, the
pressure-jump errors in panel (d) are much more tightly concentrated around
zero for the mollified--sharp model. Its largest remaining jump error is
associated with the underpredicted shock in row (d) of
Fig.~\ref{fig:model-cp-p-five-row-test}. The long negative tail of the standard
model indicates that its smeared shocks can substantially underpredict the
postprocessed jump magnitude.

\begin{figure}[!t]
  \centering
  \includegraphics[width=\textwidth]{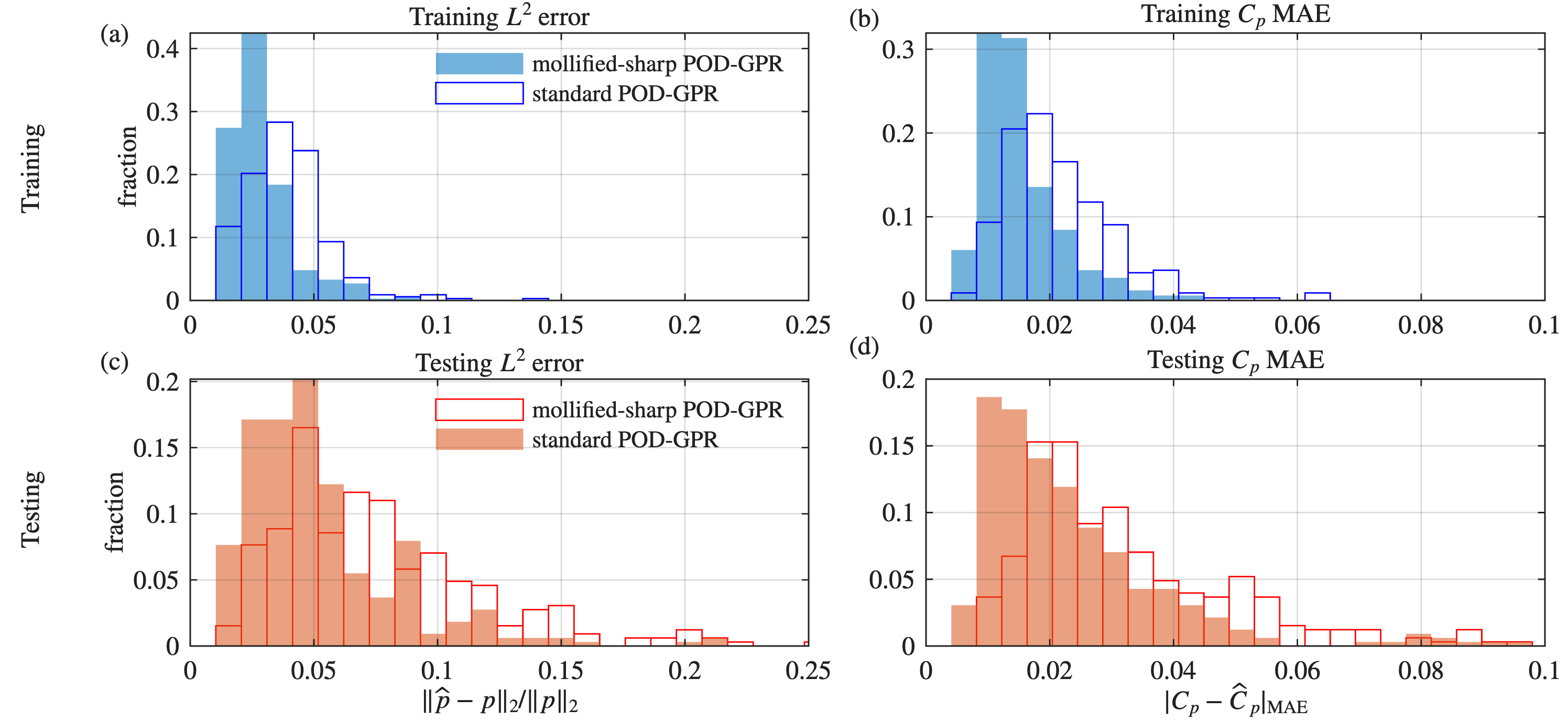}
  \caption{Distributions of model errors for the
  mollified--sharp POD--GPR model and the standard POD--GPR baseline, separated
  by sample type. Panels (a) and (b) show the training distributions of the
  full-field relative $L^2$ pressure error and airfoil-surface $C_p$ mean
  absolute error. Panels (c) and (d) show the corresponding testing
  distributions. Histogram heights are normalized as fractions of the
  training or testing set, and model styles follow the legends in each row.
  The corresponding parameter-space error maps and detailed standard-POD
  testing cases are given in Appendix~\ref{app:stdpod-comparison}.}
  \label{fig:error-comparison-histograms}
\end{figure}

Consistent with the concentration of the shock-location and pressure-jump
errors near zero in Figs.~\ref{fig:shock-feature-comparison}(a) and (d), the
global and surface-pressure errors also shift toward smaller values for the
mollified--sharp model. Fig.~\ref{fig:error-comparison-histograms} shows these
complementary sample-wise distributions, with training and testing results
separated to distinguish reconstruction at sampled parameter values from
prediction at withheld conditions.

\begin{center}
\begin{minipage}{\textwidth}
\centering
\renewcommand{\arraystretch}{1.2}
\begin{tabular}{@{}llcc@{}}
\toprule
\textbf{Sample set} & \textbf{Method} & \textbf{Mean relative $L^2$ error} & \textbf{Mean $C_p$ MAE} \\
\midrule
Training & mollified--sharp POD--GPR & 0.0288 &  0.0152\\
Training & standard POD--GPR         & 0.0390 & 0.0215\\
Training & relative difference       & -26.2\% & -29.0\%\\
\addlinespace
Testing  & mollified--sharp POD--GPR & 0.0526 & 0.0234\\
Testing  & standard POD--GPR         & 0.0765 & 0.0351\\
Testing  & relative difference       & -31.2\% & -33.2\%\\
\bottomrule
\end{tabular}
\par\medskip
\captionof{table}{Mean errors for the mollified--sharp POD--GPR
model and the standard POD--GPR baseline, separated by training and testing
samples. Absolute errors are reported as raw values. The relative-difference
rows report
$(E_{\mathrm{MS}}-E_{\mathrm{std}})/E_{\mathrm{std}}\times 100\%$, so negative
values indicate lower error for the mollified--sharp model.}
\label{tab:average-errors}
\end{minipage}
\end{center}
\FloatBarrier

The shift occurs in both metrics and for both sample sets. The testing
distributions are broader than the training distributions, as expected for
prediction at withheld conditions, and include a small number of difficult
cases such as those in Fig.~\ref{fig:model-cp-p-five-row-test}. Nevertheless,
the mean testing errors decrease by 31.2\% in the relative $L^2$ metric and by
33.2\% in the surface-$C_p$ metric. The corresponding training reductions are
26.2\% and 29.0\%, respectively.

The parameter-space maps in Fig.~\ref{fig:error-comparison} show that the
improvement is widespread but not uniform: isolated samples favor the
standard model, while the mollified--sharp representation reduces the error
over much of the shock-bearing parameter range. Together with the
shock-feature results of Fig.~\ref{fig:shock-feature-comparison}, this
comparison shows that the aggregate error reduction is accompanied by more
accurate shock placement and pressure-jump reconstruction, rather than arising
only from changes in the smooth background field.

\FloatBarrier
\section{Discussion}
\label{sec:discussion}

The results show that the probabilistic construction provides a practical way
to regularize the parameter dependence introduced by moving shocks. Assigning
an artificial distribution to a deterministic shock position and taking the
expectation replaces the localized pressure change by a smooth transition of
prescribed characteristic width. The resulting mollified field is simpler to
represent and regress over the parameter space, while the local sharp
correction retains the information needed to recover the shock. In contrast,
the standard POD--GPR model often reproduces the background pressure field
while spreading or displacing the shock-induced change. The aggregate errors
and the shock-location and pressure-jump distributions in
Fig.~\ref{fig:shock-feature-comparison} show that the improvement is associated
with more accurate reconstruction of the moving sharp features rather than
only with changes in the smooth background field.

The general construction consists of mollifying each localized pressure
change, retaining the exact residual as a local sharp correction, and
representing that correction in aligned coordinates. The detector, smoothing
kernel, geometric partition, alignment procedure, and regression method are
implementation choices. For example, a detector that gives a consistent
finite-width response could be normalized directly to construct $f_{\ell}$
and $\eta_{\ell}$, without first extracting a shock-line trace. The separate
representations require additional sharp-correction and frame regressions; in
the present implementation, the median online time was 2.24 times that of the
standard model. The regression counts, compute times, and standard-POD rank
study are reported in Appendix~\ref{app:cost}.

Because the construction introduces shock-location probability densities,
optimal transport provides a natural alternative for interpolation between
parameter points. Transporting a location density would describe the center of
the mollified transition, but would not by itself provide the signed pressure
correction, its amplitude and shape, or the presence of several independently
moving components. The present method instead regresses the position,
orientation, and scale of each local frame and represents the remaining
correction in aligned coordinates. Optimal transport could nevertheless be
used to interpolate the location densities or combined with the present
alignment procedure \citep{ehrlacher2020nonlinear,cucchiara2024model}. For the
deterministic steady snapshots studied here, the location distribution is
artificial. For genuinely uncertain or periodic shock motion, such as
transonic shock buffet \citep{moise2024connecting}, it could instead be
estimated from an ensemble or formed as a phase or time average. The
oscillation frequency, shock excursion, and other statistics could then also
be regressed, provided that their joint dependence with the pressure change
and background field is retained.

The decomposition is not limited to compression shocks or to pressure. It can
be applied to another parameter-dependent field containing a detectable
localized transition with a signed change, including the variables that jump
across contact discontinuities or material interfaces in multiphase flow. An
extension from two-dimensional shock lines to three-dimensional shock or
interface surfaces would require surface-based local coordinates and, for
strongly curved or branching structures, multiple local patches rather than a
single affine frame.

The present numerical tests are limited to pressure fields for one airfoil,
one Reynolds number, and the two parameters $(\alpha,M_\infty)$. The
reconstruction does not enforce the governing equations or the
Rankine--Hugoniot conditions, and its largest errors occur in sparsely sampled
neighborhoods close to unsteady or non-converged regimes of the source
database. Adequate parameter coverage and reliable reference solutions
therefore remain important. The fixed partition permits at most one retained
component per sector, while the principal-axis transform accounts only for
translation, rotation, and scale. More complicated changes in topology would
require component association or more general local coordinates.

\FloatBarrier
\section{Conclusion}
\label{sec:conclusion}

This paper introduced a mollified--sharp decomposition for nonintrusive
reduced-order modeling of parameterized pressure fields containing moving
shocks. An artificial probability distribution centered at each detected
shock location is used to replace the localized pressure change by a smooth
finite-width ramp in a global mollified field. The exact residual is separated
into local components, expressed in shock-aligned coordinates, and represented
independently. Separate regressions provide the global and local POD
coefficients and determine the presence, position, orientation, and scale of
each sharp correction before the components are returned to the physical
domain.

The method was evaluated on 332 training and 327 testing snapshots from the
transonic RAE2822 pressure database and compared with a standard POD--GPR
model using the same data, grid, POD-energy criterion, and regression
procedure. The mollified--sharp model reduced the mean testing relative
$L^2$ pressure error by 31.2\% and the mean testing surface-$C_p$ mean absolute
error by 33.2\%. Its shock-location and pressure-jump errors were also
more tightly concentrated near zero, and the selected predictions demonstrate
that the reconstruction accommodates multiple shock components as they move,
appear, and disappear across the parameter space.

The central representation strategy is to regularize the moving sharp
structure in the global field, retain the removed information as a local
correction, and regress its geometry separately from its aligned shape. The
detector, smoothing kernel, local partition, and regression
method can be adapted to the application. This flexibility provides a direct
path toward unsteady or uncertain shock motion, other flow variables and
moving interfaces, and higher-dimensional configurations while retaining an
explicit and nonintrusive online reconstruction.

\section*{Acknowledgments}

The author would like to acknowledge the Multidisciplinary Science and
Technology Center of the Aerospace Systems Directorate, Air Force Research
Laboratory for funding and supporting this effort through the Collaborative
Center for Design and Research of Interdisciplinary Systems program.
Distribution A: Approved for public release; distribution is unlimited.
AFRL-2026-3564.

\section*{CRediT authorship contribution statement}

Oliver T. Schmidt: Conceptualization, Methodology, Software, Validation, Formal analysis, Investigation, Data curation, Visualization, Writing -- original draft, Writing -- review and editing.

\section*{Declaration of competing interest}

The author declares no known competing financial interests or personal relationships that could have appeared to influence the work reported in this paper.

\section*{Data availability}

The RAE2822 airfoil database analyzed in this study was introduced by
\citet{catalani2023comparative} and is archived at
\url{https://zenodo.org/records/12700680}. The data are also distributed through
\url{https://huggingface.co/datasets/giovannicatalani/RAE2822_Airfoil_Dataset}.
The results reported in this article were generated using the author's original
MATLAB research code. A refactored, reorganized, and documented version of that
code is publicly available at
\url{https://github.com/olivertschmidt/mollified-sharp-transonic-airfoil}. The
public version was prepared by OpenAI Codex under the author's direction to
improve accessibility, usability, and maintainability; it automatically
downloads the required airfoil data from the Hugging Face repository.

\section*{Declaration of generative AI and AI-assisted technologies in the manuscript preparation process}

During the preparation of this work, the author used OpenAI Codex, under detailed author direction, to assist with MATLAB implementation and debugging, analysis and plotting routines, wording, and mathematical typesetting, and to refactor and document the public MATLAB release. The author takes full responsibility for the content of the publication.

\FloatBarrier
\appendix

\section*{Appendices}
\addcontentsline{toc}{section}{Appendices}

\FloatBarrier
\section{Detailed comparison with standard POD--GPR}
\label{app:stdpod-comparison}

\begin{figure}[p]
  \centering
  \includegraphics[width=\textwidth]{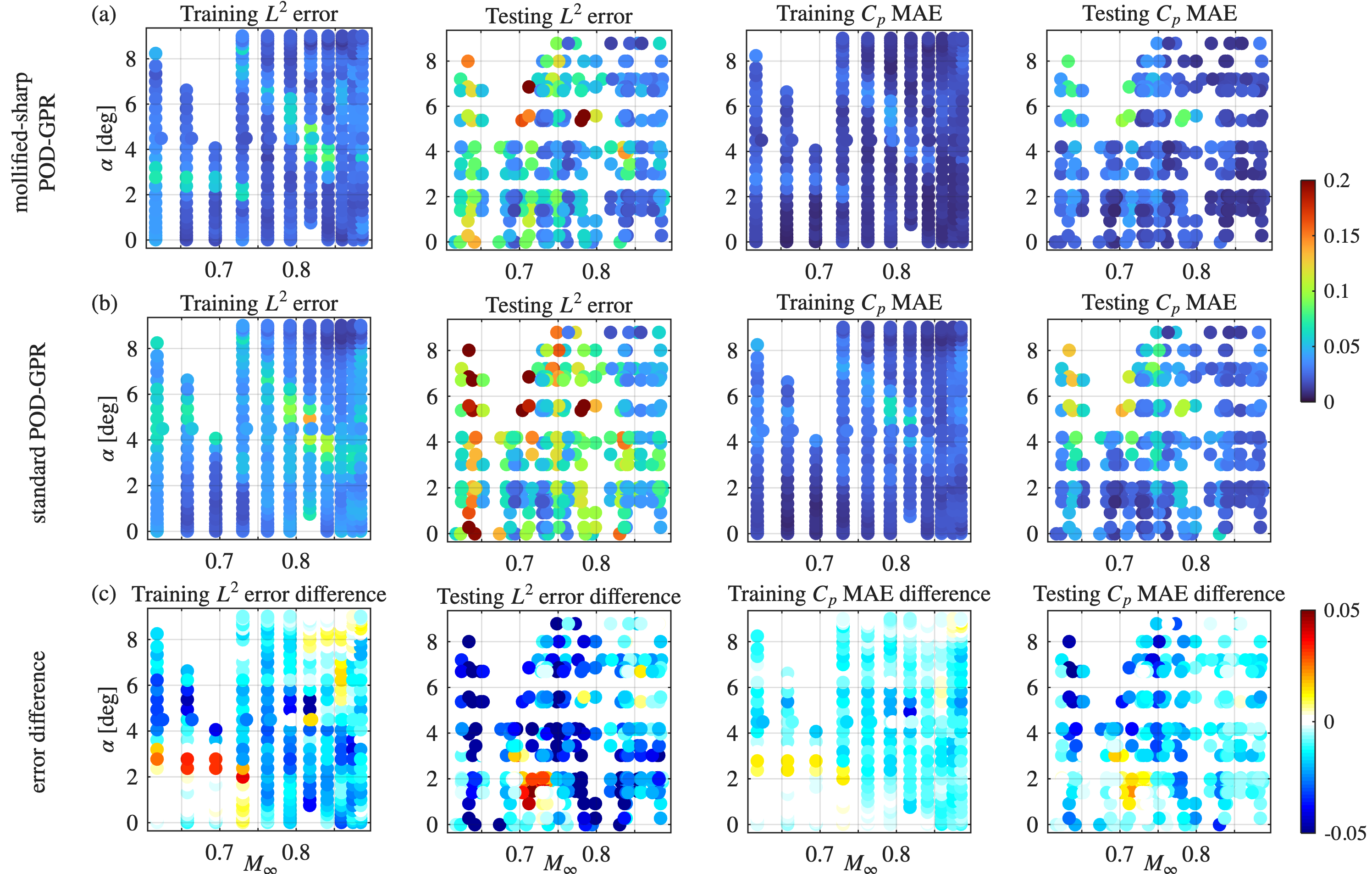}
  \caption{Error comparison between the mollified--sharp POD--GPR model and
  the standard POD--GPR baseline over the sampled parameter space. From left
  to right, the columns show the training and testing full-field relative
  $L^2$ pressure errors and the training and testing airfoil-surface $C_p$
  mean absolute errors. Rows (a) and (b) show the mollified--sharp and
  standard-model errors, respectively. Row (c) shows
  $E_{\mathrm{MS}}-E_{\mathrm{std}}$ at each common sample: negative values
  indicate lower error for the mollified--sharp model, and positive values
  indicate lower error for the standard model.}
  \label{fig:error-comparison}
\end{figure}

\begin{figure}[p]
  \centering
  \includegraphics[width=\textwidth]{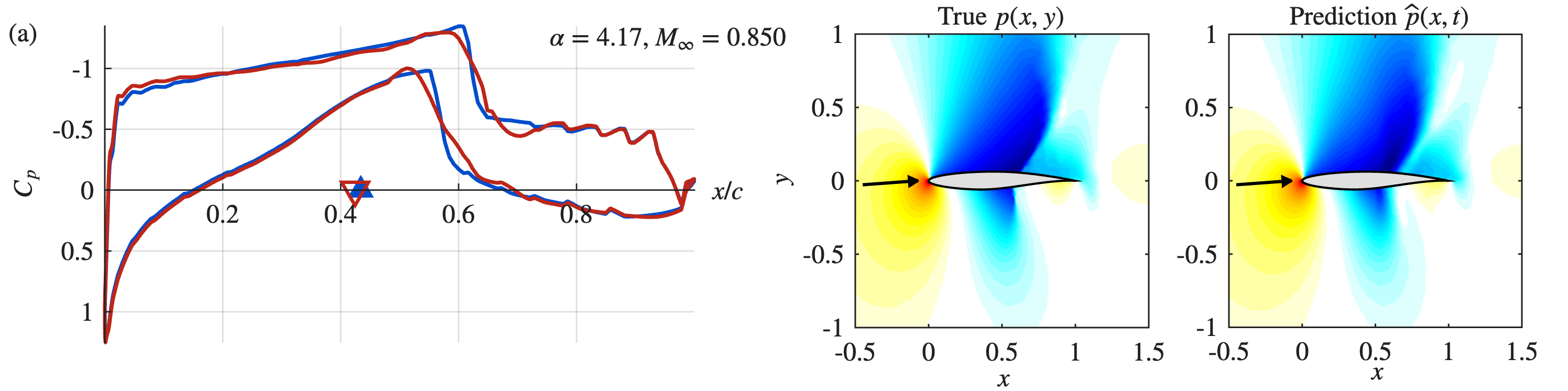}\\[-1mm]
  \includegraphics[width=\textwidth]{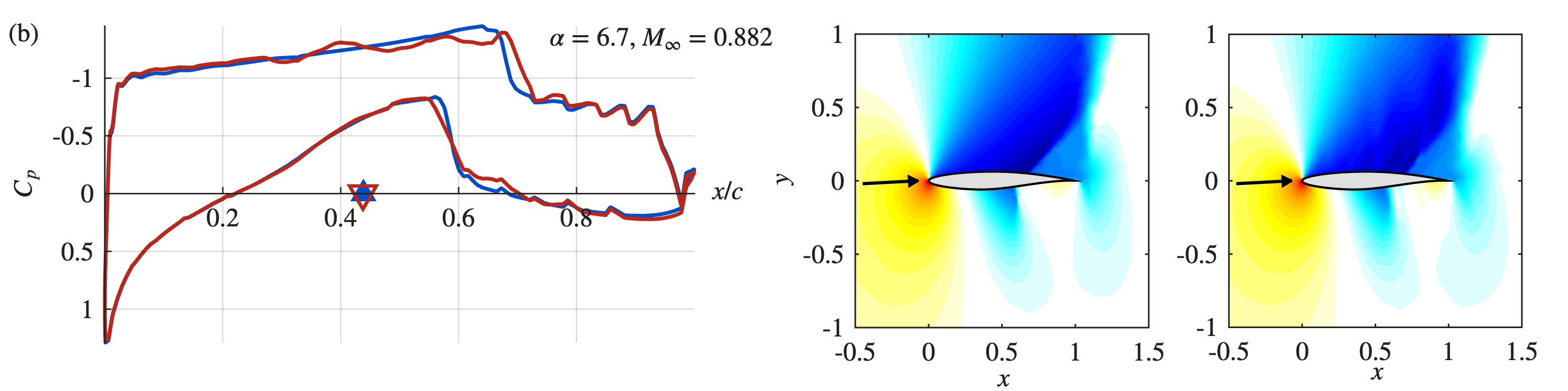}\\[-1mm]
  \includegraphics[width=\textwidth]{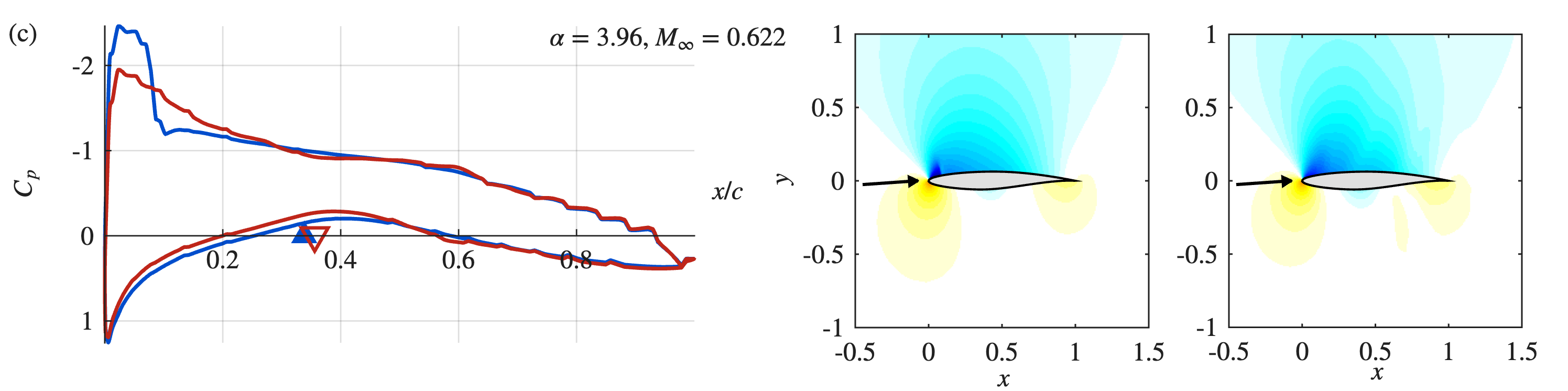}\\[-1mm]
  \includegraphics[width=\textwidth]{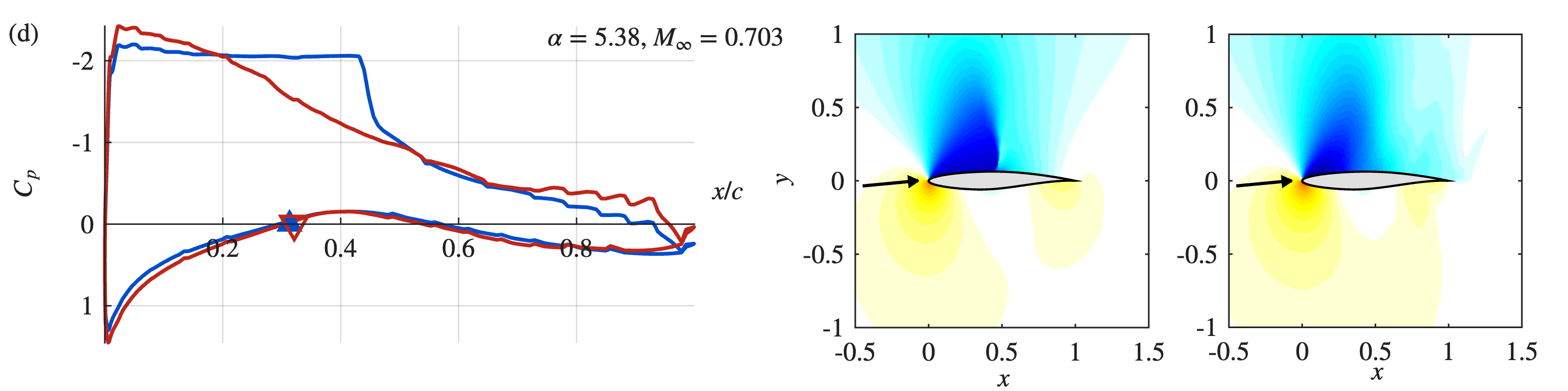}\\[-1mm]
  \includegraphics[width=\textwidth]{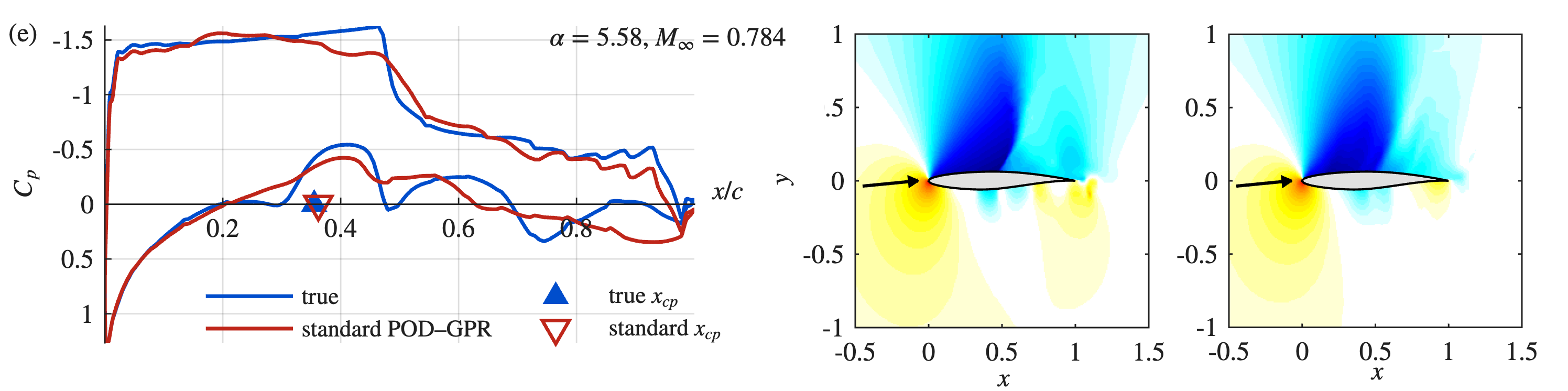}
  \caption{Standard POD--GPR predictions for the five testing cases
  shown in Fig.~\ref{fig:model-cp-p-five-row-test}, using the same samples and
  row ordering. In the left column, blue denotes the reference surface
  pressure coefficient and red the standard-POD prediction; the corresponding
  filled blue and open red triangles mark the postprocessed shock locations.
  The center and right columns show the reference and predicted pressure
  fields, respectively.}
  \label{fig:model-cp-p-stdpod-five-row}
\end{figure}

This appendix supplements the aggregate comparison in
Section~\ref{sec:results:standard-pod-comparison}. Fig.~\ref{fig:error-comparison}
locates the errors of both models in the sampled parameter space, while
Fig.~\ref{fig:model-cp-p-stdpod-five-row} shows the standard-POD predictions
for the same five testing cases examined in
Fig.~\ref{fig:model-cp-p-five-row-test}. The data partitions, pressure fields,
and error definitions are identical for the two models.

The difference maps in row (c) show that the improvement is distributed over
the shock-bearing parameter range rather than confined to a small cluster.
The testing differences are less regular than the training differences, as
expected for prediction at independently sampled parameter values. Some
isolated testing points favor the standard model.

Fig.~\ref{fig:model-cp-p-stdpod-five-row} provides the corresponding
case-by-case comparison. Each row uses the same reference snapshot and
parameter point as the equally labeled row of
Fig.~\ref{fig:model-cp-p-five-row-test}; only the reduced representation is
changed.

The standard model reproduces much of the slowly varying pressure field, but
its direct global representation tends to spread or displace the localized
shock-induced changes. In row (d), for example, the predicted upper-surface
profile does not merely place the shock at the wrong location; it replaces
the abrupt pressure change by a smooth variation with no distinct shock.
Although an individual POD--GPR prediction cannot separate basis-truncation
error from error in the regressed modal coefficients, this behavior is
consistent with the slow Kolmogorov $n$-width decay of solution sets containing
translated discontinuities. A truncated POD expansion in a fixed linear space
combines shock-bearing structures associated with different locations and can
therefore smear a jump until it is no longer resolved as a distinct feature
\citep{ohlberger2016reduced,peherstorfer2022breaking}. This distinction is
particularly clear when the surface profiles are compared with the
mollified--sharp reconstructions in
Fig.~\ref{fig:model-cp-p-five-row-test}. The comparison is qualitative at the
individual-sample level; the shock-feature and aggregate error statistics in
Figs.~\ref{fig:shock-feature-comparison} and~\ref{fig:error-comparison-histograms}
quantify the same behavior over the
complete testing set.

\subsection{Rank sensitivity and computational cost}
\label{app:cost}

The standard POD--GPR model was evaluated at retained ranks
$r=20$, 40, 59, 80, 100, 125, and 150. Increasing the rank from 59 to 150
changed the mean testing relative $L^2$ error from 0.07653 to 0.07612, a
reduction of 0.53\%, and the mean surface-$C_p$ error from 0.03505 to 0.03502,
a reduction of 0.11\%. The corresponding 90th-percentile errors showed no
consistent improvement.

\begin{center}
\begin{minipage}{\textwidth}
\centering
\renewcommand{\arraystretch}{1.15}
\begin{tabular}{@{}lrr@{}}
\toprule
\textbf{Quantity} & \textbf{Standard POD--GPR} & \textbf{Mollified--sharp POD--GPR} \\
\midrule
Global modal-coefficient GPRs     & 59      & 28 \\
Sharp-correction modal GPRs       & 0       & 271 \\
Local-frame regressions           & 0       & 24 \\
Shock-presence regressions        & 0       & 4 \\
\addlinespace
Total number of regressions       & 59      & 327 \\
Construction time                 & 30.1~s  & 54.9~s \\
Median online time per field      & 8.65~ms & 19.4~ms \\
90th-percentile online time       & 10.5~ms & 26.3~ms \\
\bottomrule
\end{tabular}
\par\medskip
\captionof{table}{Regression counts and measured construction and online
evaluation times. The construction times exclude raw data loading and the
Cartesian pressure interpolation common to both models.}
\label{tab:computational-cost}
\end{minipage}
\end{center}

The timing measurements were obtained with MATLAB R2025a Update~1 on an Apple
M1 Max using 10 computational threads. After warm-up evaluations, 240 online
times were collected for each model at 48 testing conditions with five
repetitions per condition. Every mollified--sharp prediction evaluates the 28
global modal-coefficient GPRs and four shock-presence GPRs. For each predicted
shock, it additionally evaluates the six frame regressions and the
sector-specific sharp-correction modal GPRs. The median online time was
6.32~ms with no predicted shock, 19.1~ms with one, 25.8~ms with two, and
28.6~ms with three. The measurements characterize the present implementation
rather than hardware-independent complexity.

\FloatBarrier
\section{Database-specific implementation details}
\label{app:implementation}

The main construction requires a localized transition indicator, a
finite-width unit-area probability density, a compatible modified gradient,
and a local frame for each sharp correction. These are the general ingredients of the
mollified--sharp strategy. At the level of the formulation, they are expressed
by the exact split in Eq.~\eqref{eq:decomp}, the compatible mollification in
Eqs.~\eqref{eq:mod_grad} and~\eqref{eq:p_tilde}, the additive local corrections
and their alignment in Eqs.~\eqref{eq:pstar_components}--\eqref{eq:xi}, and the
online recombination in Eq.~\eqref{eq:p_hat}. The detector, sector partition,
graph-delta trace, kernel and smoothing width, interpolation grids, numerical
thresholds, and regression choices documented below constitute one
reproducible realization for the RAE2822 database; they are not defining
requirements of the method.

\subsection{Shock indicator and shock presence}
\label{app:detection}

The mollification requires the location of each sharp transition, but it does
not require a particular shock detector. The detector described here is a
database-specific choice designed to provide a consistent front across the
reported snapshots. \citet{catalani2023comparative} used the same spatially
nonuniform hybrid CFD mesh at every operating condition. As a shock moves
through that mesh, the local sampling available to represent its pressure rise
therefore changes. Those authors also used cubic interpolation of the CFD
fields to uniform Cartesian grids for their convolutional model. The present
work instead performs a separate natural-neighbor interpolation from the
source CFD data to the Cartesian grid specified in
Appendix~\ref{app:algorithm}. Both the local source-mesh spacing and this grid
transfer can affect the apparent thickness and peak gradient of a numerically
resolved shock. The indicator and database-wide calibration below were
selected to give stable traces under these ordinary discretization effects.
They are not proposed as a universal detector, and another suitable indicator
could be substituted without changing the mollification. The chosen detector
combines an adverse pressure gradient measured along the local flow direction
with a smooth local-Mach weighting:
\begin{equation}
  s^{*}(\xvec) \;=\;
  \underbrace{\frac{\sqrt{dx\,dy}\,\max\!\bigl(\eu(\xvec) \cdot \nabla p(\xvec),\, 0\bigr)}{\Delta p_{99\%}}}_{\text{scaled adverse pressure-gradient factor}}
  \;\cdot\;
  \underbrace{\Bigl(\tfrac{1}{2} + \tfrac{1}{2}\tanh\!\bigl(2(M(\xvec) - 0.99)\bigr)\Bigr)}_{\text{local-Mach weighting factor}},
  \label{eq:s_raw}
\end{equation}
Here, $\eu=\bm{u}/\|\bm{u}\|$ is the unit vector in the local flow direction
and $M$ is the local Mach number. The positive part of
$\eu\cdot\nabla p$ selects pressure rises in the flow direction, as expected
across a compression shock. Multiplication by the characteristic grid spacing
$\sqrt{dx\,dy}$ and division by the robust, snapshot-specific pressure range
$\Delta p_{99\%}=p_{99\%}-p_{1\%}$ make this factor dimensionless and reduce
its sensitivity to isolated pressure extrema. The Mach factor equals one half
at $M=0.99$ and smoothly suppresses pressure gradients in lower-Mach regions;
it is a weighting rather than a hard sonic cutoff.

The raw score is converted to the calibrated indicator $s\in[0,1]$ using one
pair of thresholds for the entire training database:
\begin{equation}
  s(\xvec) \;=\;
  \begin{cases}
    0, & s^{*}(\xvec) < s_0,\\[2pt]
    s^{*}(\xvec)/s_1, & s_0 \le s^{*}(\xvec) \le s_1,\\[2pt]
    1, & s^{*}(\xvec) > s_1,
  \end{cases}
  \quad
  s_0 = \tfrac{1}{16}\max_{\xvec,\,k}\, s^{*}_k(\xvec),
  \quad
  s_1 = \tfrac{1}{2}\max_{\xvec,\,k}\, s^{*}_k(\xvec),
  \label{eq:s_normalize}
\end{equation}
Values below $s_0$ are removed, values above $s_1$ saturate, and the
intermediate values retain their relative amplitudes. Because $s_0/s_1=0.125$,
the smallest nonzero calibrated value is consistent with the pointwise
threshold used below. The database-wide scaling permits the same presence
criteria to be applied to every snapshot.

Let $\Omega_{\ell}$ denote the geometric sector associated with shock index
$\ell$. A shock is declared present only when the calibrated field has both a
sufficiently large pointwise value and sufficient integrated support:
\begin{equation}
  \chi_{\ell,k}
  =
  \begin{cases}
    1, & \displaystyle
    \max_{\xvec\in\Omega_{\ell}} s_k(\xvec) > \tau_s
    \ \text{and}\
    \int_{\Omega_{\ell}} s_k(\xvec)\,d\xvec > \tau_I,\\[4pt]
    0, & \text{otherwise},
  \end{cases}
  \label{eq:shock_presence_indicator}
\end{equation}
The reported results use $\tau_s=0.125$ and $\tau_I=10^{-4}$. The integral
criterion rejects isolated grid-scale responses that pass the pointwise
threshold but lack the spatial support of a resolved shock trace. If
$\chi_{\ell,k}=0$, the indicator, smoothing field, and sharp correction
associated with that sector are all set to zero for snapshot $k$.
Eq.~\eqref{eq:shock_presence_indicator} defines a binary indicator value for
each training snapshot. At prediction time, the GPR fitted to these values is
clipped to $[0,1]$ and the sector is included when its value is at least
$0.5$. This hard threshold is the simplest presence criterion used for the
present demonstration; a different regression or threshold could be
substituted without altering the decomposition.

\subsection{Graph-delta trace, probability density, and mollification weight}
\label{app:graph-delta}

For the reported implementation, the finite-width indicator is reduced to a
single streamwise location on each fixed-$y$ line. This collapse prevents the
variable numerical thickness of the detector response from becoming an
additional smoothing scale. It is an implementation choice rather than a
requirement: a detector with sufficiently consistent width and amplitude
could instead be normalized linewise and used directly to construct
$f_{\ell}$ and $\eta_{\ell}$. Let $s_{\ell,k}$ denote the calibrated indicator
restricted to sector $\ell$ in snapshot $k$:
\begin{equation}
  s_{\ell,k}(\xvec)
  =
  \begin{cases}
    s_k(\xvec), & \xvec\in\Omega_{\ell}\ \text{and}\ \chi_{\ell,k}=1,\\[2pt]
    0, & \text{otherwise}.
  \end{cases}
  \label{eq:sector_restricted_indicator}
\end{equation}
On every line containing nonzero indicator values, their weighted streamwise
centroid defines the shock coordinate:
\begin{equation}
  \xstar_{\ell,k}(y) \;=\; \frac{\sum_{x'} x'\,s_{\ell,k}(x',y)}{\sum_{x'} s_{\ell,k}(x',y)},
  \label{eq:xstar}
\end{equation}
The implementation assigns this coordinate to the nearest active grid node.
This converts a several-cell-wide detector response into a one-node trace
while preserving its detected center to grid resolution. The corresponding continuous
representation is the graph delta
\begin{equation}
  \dstar_{\ell,k}(\xvec)
  =
  \begin{cases}
    \delta\!\bigl(x-\xstar_{\ell,k}(y)\bigr),
      & \chi_{\ell,k}=1\ \text{and}\ \sum_{x'}s_{\ell,k}(x',y)>0,\\[2pt]
    0, & \text{otherwise}.
  \end{cases}
  \label{eq:dstar_definition}
\end{equation}
which carries one unit of mass in $x$ on every active line.

On the Cartesian grid, Eq.~\eqref{eq:dstar_definition} is represented by one
unit-valued node per active line. The reported implementation uses the
two-dimensional radial kernel $\fsig^{(2)}$ in Eq.~\eqref{eq:fl}, so the
convolution also couples neighboring $y$ lines. The discrete kernel is
normalized to unit total mass, and the smoothed field is renormalized where
its support intersects the outer boundary of the Cartesian domain. For the
upper- and lower-airfoil sectors, the source trace and smoothed field are
restricted to the corresponding side of the centerline. A nearest-neighbor
continuation from the closest detected trace toward the centerline prevents
the near-wall support from becoming disconnected where the airfoil interrupts
the trace.

The convolution first supplies a preliminary nonnegative profile. Its
linewise peak normalization gives the preliminary mollification weight
$\eta'_{\ell}$. For an isolated component, normalization of the same profile
by its line integral gives the unit-area probability density $f_{\ell}$; the
two normalizations then differ only by a scalar and preserve the same center,
support, and shape. If preliminary weights from different sectors overlap,
they are first partitioned proportionally so that their sum does not exceed
one. Each partitioned $\eta_{\ell}$ is then normalized by its line integral
over the active support interval to obtain the final $f_{\ell}$. Thus, in an
overlap region, the final linewise probability density is derived from the
partitioned weight rather than directly from the unconstrained convolution.
The exact discrete construction is given below in
Eqs.~\eqref{eq:algorithm_eta_partition}
and~\eqref{eq:algorithm_line_pdf_jump}.

The graph-delta convention also explains how shock obliquity changes the
effective streamwise smoothing width. Unlike an arc-length delta on the front,
$\dstar_{\ell}$ carries unit mass per fixed-$y$ line. Denote by
$\fsig^{(1)}$ the streamwise kernel profile induced by the two-dimensional
convolution. For a straight front whose tangent makes an angle $\theta$ with
the vertical direction normal to the streamwise sampling line,
\[
  f_{\ell}(x,y)
  =
  \cos\theta\,\fsig^{(1)}\!\bigl((x-\xstar_{\ell}(y))\cos\theta\bigr).
\]
The profile remains centered at $\xstar_{\ell}(y)$ but is stretched in the
streamwise direction to the width $\sigma/\cos\theta$. At normal incidence it
reduces to
$f_{\ell}(x,y)=\fsig^{(1)}\!\bigl(x-\xstar_{\ell}(y)\bigr)$, with width
$\sigma$; front curvature can introduce some linewise asymmetry.

The detector thresholds determine which sectors and lines are active and
where their graph traces are placed. Once a trace has been selected, the
kernel fixes the nominal smoothing width, while the pressure data determine
the signed amplitude through Eq.~\eqref{eq:jump}. Thus detector calibration,
smoothing width, and pressure-jump estimation play distinct roles, even though
all three ultimately affect the decomposed fields used to construct the
reduced-order model.

\FloatBarrier
\subsection{Algorithmic realization}
\label{app:algorithm}

This subsection collects the numerical choices needed to reproduce the
reported model. Dimensional source coordinates are converted to chord
coordinates by multiplication with $V_\infty/(Re\,\nu)$, where
$V_\infty=M_\infty a_\infty$, $Re=6.5\times10^6$, and
$\nu=1.46\times10^{-5}~\mathrm{m^2\,s^{-1}}$. Pressure, Mach number, and
velocity components are transferred to a uniform $250\times250$ Cartesian grid
covering $x\in[-0.5,1.5]$ and $y\in[-1,1]$ by natural-neighbor interpolation,
with nearest-neighbor extrapolation. Spatial derivatives are finite-difference
approximations, and all streamwise line integrals are evaluated by the
trapezoidal rule at fixed $y_j$.

\subsubsection*{Training}

\begin{enumerate}
  \item \textbf{Grid fields, shock detection, and graph traces.}
  For every retained training parameter
  $\muvec_k=(\alpha_k,M_{\infty,k})$, interpolate pressure, local Mach number,
  and both velocity components to the Cartesian grid. Samples containing
  nonfinite source data are discarded. Before differentiation and
  reconstruction, pressure and flow values at grid points inside the airfoil
  are filled from the nearest exterior point.

  Evaluate the shock indicator using Eqs.~\eqref{eq:s_raw}
  and~\eqref{eq:s_normalize}. The upstream region is excluded, leaving the four
  sectors $\ell\in\{\sct{I},\sct{II},\sct{III},\sct{IV}\}$ shown in
  Fig.~\ref{fig:sector-definitions}, with at most one retained shock per
  sector. Eq.~\eqref{eq:shock_presence_indicator} supplies
  $\chi_{\ell,k}$. For mollification, replace the finite-width response on
  every active fixed-$y_j$ line by a unit-valued node at the nearest active
  grid point to the weighted centroid in Eq.~\eqref{eq:xstar}. This is the
  discrete graph trace of Eq.~\eqref{eq:dstar_definition}. The uncollapsed,
  thresholded indicator remains available for detection and alignment
  diagnostics.

  \item \textbf{Smoothing weights and component partition.}
  Restrict each active trace to its sector and, for the airfoil sectors, to
  the corresponding side of the centerline. Empty same-side cells between the
  closest detected trace and the centerline are continued by nearest neighbor
  before smoothing. This bridge keeps the support connected near the airfoil
  wall.

  Convolve each trace with the compact radial Wendland $C^4$ kernel. In terms
  of the radius normalized by its support radius, its unnormalized profile is
  $(1-r)^6_+(35r^2+18r+3)$. The sampled kernel is normalized to unit discrete
  mass, and its support radius is chosen so that its radial second moment
  equals that of a two-dimensional Gaussian with
  $\sigma=0.075$ in chord-normalized coordinates. The resulting Wendland
  support radius is approximately $0.326$ chord lengths. Normalize the smoothed
  field by its streamwise maximum on each fixed-$y_j$ line to obtain the preliminary weight
  $\eta'_{\ell}$. Where preliminary component weights overlap, partition them
  according to
  \begin{equation}
    \eta(\xvec)=\min\!\left(1,\sum_m \eta'_m(\xvec)\right),
    \qquad
    \eta_{\ell}(\xvec)
    =
    \begin{cases}
      \eta'_{\ell}(\xvec)\,\eta(\xvec)\big/\sum_m\eta'_m(\xvec),
        & \sum_m\eta'_m(\xvec)>0,\\[2pt]
      0, & \sum_m\eta'_m(\xvec)=0,
    \end{cases}
    \label{eq:algorithm_eta_partition}
  \end{equation}
  so that $\sum_{\ell}\eta_{\ell}=\eta\leq1$. If no sector contains a
  detected shock, the decomposition is assigned directly as $\tp_k=p_k$ and
  $\pstar_k=0$.

  \item \textbf{Local supports and modified gradients.}
  On a fixed-$y_j$ line, mark nodes for which at least one $\eta_{\ell}$
  exceeds the support tolerance $10^{-3}$. Split the marked nodes into
  contiguous intervals $[x_-,x_+]$ and process each interval independently.
  For every active component with positive line integral, form the unit-area
  linewise probability density and signed pressure change
  \begin{equation}
    f_{\ell}(x,y_j)=
      \frac{\eta_{\ell}(x,y_j)}
           {\displaystyle\int_{x_-}^{x_+}\eta_{\ell}(x',y_j)\,dx'},
    \qquad
    \Delta p_{\ell}(y_j)=
      \int_{x_-}^{x_+}\eta_{\ell}(x,y_j)
      \frac{\partial p_k}{\partial x}(x,y_j)\,dx.
    \label{eq:algorithm_line_pdf_jump}
  \end{equation}
  The modified gradient is then the discrete counterpart of
  Eq.~\eqref{eq:mod_grad}:
  \begin{equation}
    \frac{\partial \tp_k}{\partial x}
    =
    (1-\eta)\frac{\partial p_k}{\partial x}
    +
    \sum_{\ell}\Delta p_{\ell}(y_j)f_{\ell},
    \label{eq:algorithm_target_gradient}
  \end{equation}
  Terms belonging to inactive components are omitted. Outside the active
  intervals, retain $\tp_k=p_k$.

  \item \textbf{Pressure reconstruction and additive sharp components.}
  Integrate Eq.~\eqref{eq:algorithm_target_gradient} from $x_-$ using
  $p_k(x_-,y_j)$ as the integration constant. A linear endpoint correction is
  then added over the interval so that the reconstructed pressure also equals
  $p_k(x_+,y_j)$ at the right endpoint. This correction removes the small
  finite-difference and quadrature imbalance that would otherwise introduce a
  pressure offset outside the mollification support.

  Initialize the component corrections on the same interval by integrating
  the deficit gradients
  \begin{equation}
    \frac{\partial \pstar_{\ell,k}}{\partial x}
    =
    \eta_{\ell}\frac{\partial p_k}{\partial x}
    -
    \Delta p_{\ell}(y_j)f_{\ell},
    \qquad
    \pstar_{\ell,k}(x_-,y_j)=0.
    \label{eq:algorithm_pstar_init}
  \end{equation}
  After reconstructing $\tp_k$, compute the total correction from the exact
  residual $\pstar_k=p_k-\tp_k$. Distribute any residual difference between
  this field and the sum of the integrated components using
  \begin{equation}
    \rho_{\ell}(\xvec)=
    \frac{\eta_{\ell}(\xvec)}
         {\sum_m\eta_m(\xvec)}
    \quad\text{where } \sum_m\eta_m(\xvec)>0,
    \label{eq:algorithm_partition_weights}
  \end{equation}
  wherever the denominator is positive, and set the weights to zero
  elsewhere. At each such node, add to component $\ell$ its weight
  $\rho_{\ell}$ times the difference between the exact residual $\pstar_k$ and
  the current sum of the component fields. Reapply this partition after all
  intervals have been processed so that
  $\pstar_k=\sum_{\ell}\pstar_{\ell,k}$ holds to machine precision on the
  Cartesian grid. The endpoint and partition corrections remove the
  differentiation and quadrature imbalance of the discrete construction; they
  do not change the exact total residual.

  \item \textbf{Shock alignment and local snapshots.}
  For every detected sector shock, use $\eta_{\ell,k}$ as the weight in the
  centroid and covariance calculation of Eq.~\eqref{eq:pca_cov}. Assign the
  covariance axes so that $\epar$ follows the shock tangent and points away
  from the airfoil centerline, while $\eperp$ is the orthogonal downstream
  direction. Reprojecting the $\eta_{\ell,k}$ moments onto these assigned axes
  gives $\lambda_{\parallel,\ell}$ and $\lambda_{\perp,\ell}$ in
  Eq.~\eqref{eq:pca_axes}.

  Interpolate $\pstar_{\ell,k}$ to a uniform $150\times150$ local grid on
  $(\xiperp,\xipar)\in[-1,1]^2$ using Eq.~\eqref{eq:xi} and the half-widths
  \begin{equation}
    L_{\perp,\ell}=3\sqrt{\lambda_{\perp,\ell}},
    \qquad
    L_{\parallel,\ell}=3\sqrt{\lambda_{\parallel,\ell}},
    \label{eq:algorithm_alignment_widths}
  \end{equation}
  Linear interpolation is used, with zero assigned wherever a local-grid
  point maps outside the physical interpolation domain. A detected shock is
  excluded from the local ensemble if the weight has zero integral, the
  centroid or axes are nonfinite, the tangential variance is nonpositive, or
  the mapped $\eta_{\ell}$ field has no nonzero local-grid values. At least three aligned
  snapshots are required to construct a sector model.

  \item \textbf{Shock-presence and frame regressions.}
  Fit one GPR per sector to the binary indicator values $\chi_{\ell,k}$ from all
  training samples. Its continuous output is clipped to $[0,1]$ and
  the sector is included at prediction time when this output is at least
  $0.5$. Using only the valid aligned snapshots, fit separate GPRs to the
  entries of the stored frame descriptor
  \begin{equation}
    \theta_{\ell}
    =
    \bigl(
      \bar{x}_{\ell},\bar{y}_{\ell},
      \log\lambda_{\parallel,\ell},\log\lambda_{\perp,\ell},
      \cos 2\vartheta_{\ell},\sin 2\vartheta_{\ell}
    \bigr),
    \qquad
    \epar=(\cos\vartheta_{\ell},\sin\vartheta_{\ell}),
    \label{eq:algorithm_transform_descriptor}
  \end{equation}
  This is the doubled-angle implementation of the conceptual descriptor in
  Eq.~\eqref{eq:theta_alignment}. The logarithms enforce positive predicted
  scale eigenvalues after exponentiation, while the doubled angle respects
  the sign indeterminacy of a principal axis.

  \item \textbf{POD--GPR models.}
  Form the global snapshot matrix from all $\tp_k$ fields and one
  sector-specific matrix from each valid aligned set of
  $\pstar_{\ell,k}$ fields. Subtract the corresponding snapshot mean, compute
  an economy singular-value decomposition, and retain the smallest rank
  capturing $99.9\%$ of the cumulative snapshot energy. All reported POD
  bases use the standard Euclidean inner product without additional spatial
  weights. The global basis uses the complete Cartesian arrays after the
  nearest-exterior fill inside the airfoil, and the sharp-correction bases use
  the uniform shock-aligned grid. No additional modal cap is imposed. The
  resulting representations are
  Eqs.~\eqref{eq:p_tilde_hat} and~\eqref{eq:p_star_hat_aligned}.

  Fit each nonconstant modal coefficient independently over
  $\muvec=(\alpha,M_\infty)$ in MATLAB R2025a using \texttt{fitrgp}, with an
  anisotropic squared-exponential covariance, separate length scales for
  $\alpha$ and $M_\infty$, and standardized parameter coordinates.
  Hyperparameter initialization, noise estimation, and optimization use the
  MATLAB R2025a defaults. Constant coefficients are stored directly. The same
  GPR form is used for the presence and frame quantities above. All reported
  bases and coefficients are obtained from the corresponding full fields.
\end{enumerate}

\subsubsection*{Prediction}

\begin{enumerate}
  \item \textbf{Mollified field.}
  At a requested parameter $\muvec$, evaluate the global modal-coefficient
  regressions and reconstruct $\tpr(\xvec;\muvec)$ using
  Eq.~\eqref{eq:p_tilde_hat}.

  \item \textbf{Sector presence and frame recovery.}
  Evaluate $\hat{\chi}_{\ell}(\muvec)$ for each sector. If its value is below
  $0.5$, or if fewer than three aligned snapshots were available during
  training, set the sector contribution to zero. Otherwise, predict the frame
  descriptor. Recover $\lambda_{\parallel,\ell}$ and
  $\lambda_{\perp,\ell}$ by exponentiation and compute
  $\vartheta_{\ell}=\tfrac{1}{2}\operatorname{atan2}
  (\sin2\vartheta_{\ell},\cos2\vartheta_{\ell})$. Orient $\epar$ outward from
  the centerline and $\eperp$ downstream. Reject the component if its
  predicted centroid lies outside the associated sector.

  \item \textbf{Sharp corrections and final pressure.}
  Evaluate the sector POD--GPR model to obtain
  $\pstarr_{\ell}(\xivec;\muvec)$ from
  Eq.~\eqref{eq:p_star_hat_aligned}. Use the predicted centroid, axes, and
  half-widths of Eq.~\eqref{eq:algorithm_alignment_widths} in the inverse of
  Eq.~\eqref{eq:xi} to map the aligned field to the physical grid. Assign zero
  outside the local interpolation domain and restrict upper- and
  lower-airfoil components to the corresponding side of the centerline.
  Finally, sum the accepted components according to
  Eq.~\eqref{eq:p_star_hat_combined} and add them to $\tpr$ using
  Eq.~\eqref{eq:p_hat}. Apply the same nearest-exterior fill inside the
  airfoil as during training.
\end{enumerate}

\bibliographystyle{cas-model2-names}
\bibliography{cas-refs}

\end{document}